\pdfoutput=1
\documentclass[sigconf,natbib,nonacm]{acmart} 
\usepackage{tabularray}
\usepackage{rotating}
\usepackage{listings}
\usepackage{multirow}
\usepackage{graphics}
\usepackage{graphicx}
\usepackage{adjustbox}
\usepackage{dirtytalk}
\usepackage{algorithm}
\usepackage{algpseudocode}
\usepackage{float}
\usepackage{todonotes}
\usepackage{bm}
\usepackage{enumerate}

\AtBeginDocument{%
  \providecommand\BibTeX{{%
    \normalfont B\kern-0.5em{\scshape i\kern-0.25em b}\kern-0.8em\TeX}}}

\begin{document}
\title{A Hierarchical Neural Framework for Classification and its Explanation in Large Unstructured Legal Documents}

\author{Nishchal Prasad}
\affiliation{
  \city{IRIT, Toulouse}
  \country{France}}
\email{Nishchal.Prasad@irit.fr}

\author{Mohand Boughanem}
\affiliation{
  \city{IRIT, Toulouse}
  \country{France}
}
\email{Mohand.Boughanem@irit.fr}

\author{Taoufiq Dkaki}
\affiliation{
 \city{IRIT, Toulouse}
 \country{France}}
 \email{Taoufiq.Dkaki@irit.fr}

\begin{abstract}
Automatic legal judgment prediction and its explanation suffer from the problem of long case documents exceeding tens of thousands of words, in general, and having a non-uniform structure. Predicting judgments from such documents and extracting their explanation becomes a challenging task, more so on documents with no structural annotation. We define this problem as "scarce annotated legal documents" and explore their lack of structural information and their long lengths with a deep-learning-based classification framework which we call MESc; "Multi-stage Encoder-based Supervised with-clustering”; for judgment prediction. 
We explore the adaptability of LLMs with multi-billion parameters (GPT-Neo, and GPT-J) to legal texts and their intra-domain(legal) transfer learning capacity. Alongside this, we compare their performance and adaptability with MESc and the impact of combining embeddings from their last layers. For such hierarchical models, we also propose an explanation extraction algorithm named ORSE; Occlusion sensitivity-based Relevant Sentence Extractor; based on the input-occlusion sensitivity of the model, to explain the predictions with the most relevant sentences from the document. We explore these methods and test their effectiveness with extensive experiments and ablation studies on legal documents from India, the European Union, and the United States with the ILDC dataset and a subset of the LexGLUE dataset. MESc achieves a minimum total performance gain of approximately 2 points over previous state-of-the-art proposed methods, while ORSE applied on MESc achieves a total average gain of 50\% over the baseline explainability scores.
\end{abstract}

\keywords{Extractive explanation, Legal judgment prediction, Scarce annotated documents, Long document classification, Multi-stage classification framework}
\maketitle

\section{Introduction}
A legal case proceeding cycle generally involves pre-filing and investigation, pleadings, discovery, 
pre-trial motions, trial, post-trial motions and appeals, and enforcement\footnote{\url{https://www.law.cornell.edu/wex/civil_procedure}}.
Out of the many steps in this cycle involving case filing, appeals, etc., the lawyer or the judge needs to analyze each case giving a certain time to come to a conclusion. This can involve analyzing vast amounts of data and legal precedents, which can be a time-consuming process given the complexity and length of the case. 
The number of legal cases in a country is also proportionally related to its population. 
This leads to a backlog of cases, especially in countries with huge populations, ultimately setting back the progress of its legal system \footnote{\url{https://www.globaltimes.cn/page/202204/1260044.shtml}}\cite{katju_pending_cases}.

Automating such legal case procedures can help speed up and strengthen the decision-making process, saving time and benefiting both the legal authorities and the people involved. (The scope of this work is within the ethical considerations discussed in the section \ref{ethical})

One of the fundamental problems that deal with this larger component is the prediction of the outcome 
based just on the case's raw texts (which include the facts, arguments, appeals, etc. except the final decision), which corresponds to the typical real-life scenario. 
While alongside, 
the reasoning as to why such a prediction (decision/judgment) was made is essential to understand, rely upon, and use that prediction in a more informative way.

There have been many machine learning techniques applied to legal texts in the past 
to predict the judgments as a text classification problem (\cite{survey_1}, \cite{survey_2}). While it may seem like a general text classification task, legal texts differ from general texts and are rather more complex, broadly in two ways, i.e. structure and syntax and, lexicon and grammar 
(\cite{diff_from_generalTxt_1}, \cite{legal-bert}, \cite{diff_from_generalTxt_2}). The structure of legal case documents (preamble, facts, appeals, facts, etc.) is not uniform in most settings and their complex syntax and lexicon make it more difficult to annotate, 
requiring only legal professionals. This adds to another challenge of long lengths of legal case documents, reaching more than 10000 tokens (Table \ref{dataset_des}). The lack of structure information 
and the long lengths of structure-specific legal documents can be defined as a more specific text classification problem, which in our work we call \say{scarce-annotated documents}. 

\textit{What are scarce-annotated legal documents?}
Legal documents have a structure that comprises a preamble, facts, justifications, case arguments, appeals, and old decisions. Except for the preamble (the heading), the remaining parts (with varying lengths) can occur in no specific order. This also varies with individual documents, making them non-uniform in nature. In such a case, it becomes necessary to provide labels (annotate) defining this structure for a model, to understand the relevant parts of the document and help make a better prediction. These relevant parts also get hidden as the length of the document increases (10's of thousands of tokens), hence making it noisy.
We define this scarcity of structure information in large legal case documents as scarce-annotated.
One thing to be noted here is that the documents have their class (final decision/judgment) label(s) but they don't give any other information regarding its structure. 

We explore this problem of classification of large scarce-annotated legal documents with their explanation. 

For classification, to approximate the structure we cluster the features generated (after processing individual parts of these documents) from a language model (transformer-based), and then use them alongside to make the predictions. With this, we try to see if these approximated labels help in judgment prediction with experiments on four different datasets (ILDC\cite{ildc-cjpe} and LexGLUE's \cite{lexglue} ECtHR(A), ECtHR(B) and SCOTUS) which can be found in the later sections of the paper. The work on classification can be summarised below: 
\begin{itemize}
    \item We define the problem of judgment prediction from large scarce-annotated legal documents and propose a multi-stage neural classification framework named \say{Multi-stage Encoder-based Supervised with-clustering} (MESc). This works by extracting embeddings from the last four layers of a fine-tuned encoder of a large language model (LLM) and using an unsupervised clustering mechanism on them to approximate structure labels. Alongside the embeddings, these labels are processed through another set of transformer encoder layers for final classification. 
    \item With ablation on MESc we show the importance of combining features from the last layers of transformer-based large language models (LLMs) (BERT \cite{bert}, GPT-Neo \cite{gpt-neo}, GPT-J \cite{gpt-j}) for such documents, with the impact on their prediction upon using the approximated structure labels.
    \item  We also study the adaptability of multi-billion parametered LLMs to the hierarchical framework and scarce-annotated legal documents while studying their intra-domain(legal) transfer learning capacity (both with fine-tuning and in MESc). 
    \item MESc achieves a minimum gain of $\approx$ 2 points in classification on ILDC and the LexGLUE's subset.
\end{itemize}
To explain the predictions, we developed a novel explanation extraction algorithm to rank and extract relevant sentences that impacted the prediction. 
While these sentences cannot exactly serve as the explanation as they are just marked sentences, but 
in a situation where no annotations are available to train an abstractive explanative algorithm these sentences can serve as a representative, for an explanation, to guide an expert on what led to a certain prediction. The results also showcase the challenge associated with such documents with a scope for future developments.



The work on explanation is summarized below:
\begin{itemize}
    \item We propose an extractive explanation algorithm, Occlusion sensitivity based Relevant Sentence Extractor (ORSE), based on the input sensitivity of a model 
    which ranks relevant sentences from a document, to serve as an explanation for its predicted class.
    \item We test ORSE for explanation extraction on ILDC$_{expert}$ \cite{ildc-cjpe}, with GPT-J\cite{gpt-j} and InLegalBERT\cite{InLegalBERT} obtaining new benchmarks.
     
\end{itemize}

\section{Related works}
\label{related_works}
Several strategies have been investigated in the past with machine-learning techniques to predict the result
of legal cases in specific categories (criminal, civil, etc.) with rich annotations (Xiao et al. \cite{Chaojun}, Xu et al. \cite{xu-etal}, Zhong et al. \cite{zhong-etal-2018-legal}, Chen et al. \cite{Huajie}). These studies on well-structured and annotated legal documents show the effect and importance of having good structural information. While creating such a dataset is both time and resource (highly skilled) demanding, researchers have worked on legal documents in a more general and raw setting. 
 
 Chalkidis et al. \cite{etchr_a} presented a dataset of European Court of Human Rights case proceedings in English, with each case assigned a score indicating its importance. They described a Legal Judgment Prediction (LJP) task for their dataset, which seeks to predict the outcome of a court case using the case facts and law violations. They also curated another version of this dataset \cite{ecthr_b} to give a rational explanation for the decision prediction made on them. In the US case setting, Kaufman et al. \cite{usa_kaufman_kraft} used AdaBoost decision tree to predict the U.S. Supreme Court rulings. Tuggener et al. \cite{ledgar} curated LEDGAR, a multilabel dataset of legal provisions in US contracts. Malik et al. \cite{ildc-cjpe} curated the Indian Legal Document Corpus (ILDC) of unannotated and unstructured documents, and used it to build baseline models for their Case Judgment Prediction and Explanation (CJPE) task 
 upon which Prasad et al. \cite{CIRCLE22} showed the possibility of intra-domain(legal) transfer learning using LEGAL-BERT on Indian legal texts (ILDC).

Pretrained language models based on transformers (Devlin et al.\cite{bert}, Vaswani et al.\cite{transformer}) have shown widespread success in all fields of natural language processing (NLP) but only for short texts spanning a few hundred tokens. There have been several approaches to handle longer sequences with transformer encoders (Beltagy et al.\cite{longformer}, Kitaev et al.\cite{reformer}, Zaheer et al.\cite{bigbird}, Ainslie et al. \cite{ETC_transformer}) 
demanding expensive domain-specific pretraining for adaptation to legal texts, and are not guaranteed to scale compared to their vanilla counterparts (Tay et al. \cite{Tay_Scaling_Laws}). 
Since we try to get the structure information 
of the document, we choose to process the document in short sequences rather than as a whole. These short sequences will help us to learn and approximate their structure labels. 
So, we take a different approach to handle large documents with smaller pre-trained transformer encoder models (such as BERT \cite{bert}) based on the hierarchical idea of \say{divide, learn and combine} (Chalkidis et al.\cite{HAT}, Zhang et al. \cite{hibert}, Yang et al. \cite{hierarchical-attention-networks}), where the document is split (into parts then sentences and words, etc.) and features of each component are learned and combined together hierarchically from bottom-up to get the whole document's representation. 

The domain-specific pre-training of transformer encoders has also accelerated the development of NLP in legal systems with better performance as compared to the general pre-trained variants (Chalkidis et al.\cite{legal-bert}'s LEGAL-BERT trained on court cases of the US, UK, and EU, 
Zheng et al. \cite{case-hold}'s 
BERT trained on US court cases dataset CaseHOLD, Shounak et al. \cite{InLegalBERT}'s InLegalBERT and InCaseLawBERT trained on the Indian legal cases). Recently, with the emergence of multi-billion parametered LLMs such as GPT-3 \cite{GPT-3}, LLaMA \cite{llama}, LaMDA \cite{lamda}, and their superior performance in natural language understanding, researchers have tried to adapt (with few-shot learning) their smaller variants to legal texts (Trautmann et al. \cite{gpt-legal-promt_0}, Yu et al. \cite{gpt-legal-promt_1}).
We check their adaptability with the hierarchical idea of \say{divide, learn and combine} and their intra-domain(legal) transfer-learning with full-fine tuning compared to the intra-domain pre-training (as done in LEGAL-BERT, InLegal-BERT).
To do so we use three such variants of GPT (GPT-Neo (1.3 and 2.7)\cite{gpt-neo}, GPT-J\cite{gpt-j}) pre-trained on Pile\cite{pile}, which has a subset (FreeLaw) of court opinions of US legal cases.

To rely upon the judgment prediction of a legal case an explanation leading to that judgment is of paramount importance. In scenarios where there is a lack of explanations annotation of legal texts, an extractive explanation method is a good fit to create interpretations of the predicted judgments. 
To provide interpretations for their judgment prediction task, \cite{Zhong_Wang_2020} employed Deep Reinforcement learning and created a "question-answering" based model called QAjudge. Jiang et al. \cite{jiang-etal-interpretable} try to extract readable snippets of texts (called rationale) from legal texts using reinforcement learning for their judgment classification problem. To improve the interpretability of charge prediction systems Ye et al. \cite{ye-etal-interpretable} propose a label-conditioned Seq2Seq model, which, for a given predicted charge 
chooses relevant reasoning in the legal document. Since we try to develop an extractive explanation algorithm using no training data and relying solely on a trained model we turn toward the idea of input sensitivity of a model (Zeiler et al. \cite{Zeiler}, Petsiuk et al. \cite{rise}) which has been used in the interpretation of computer vision models, where the pixels are scored (according to a scoring parameter) against their absence in the input, and finally they are chosen according to the desirability of the scores (higher or lower). 


\section{Method}



\begin{figure}[h]
  \centering
  \includegraphics[width=7cm]{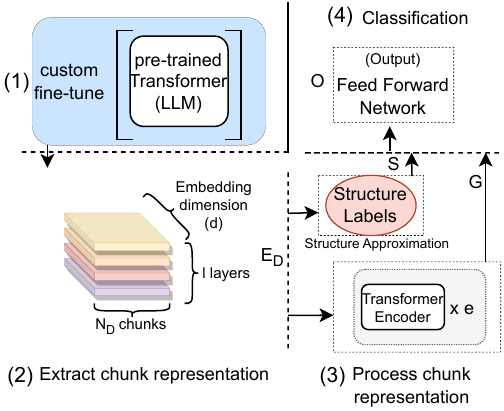}
  \caption{MESc classification Framework}
  \label{fig:proposed_model}
\end{figure}

\subsection{Classification Framework (MESc)}
To handle large documents MESc architecture shares the general hierarchical idea of divide, learn and combine (\cite{HAT}, \cite{hibert}, \cite{hierarchical-attention-networks}) but it differs from the previous works in the following main aspects: 

a) It employs custom fine-tuning and uses the last four layers of the fine-tuned transformer encoder for extracting representations for parts(chunks) of the document.

b) It approximates the document structure by applying unsupervised learning (clustering) on these representations' embeddings and uses this information alongside, for classification. %

c) Different configurations of transformer encoder layers are used and experimented with to attend over the combined embeddings to learn the intra-chunk representation to get a global document representation. 

d) Divide the process into four stages, custom fine-tuning, extracting embeddings, processing the embeddings (supervised + unsupervised learning), and classification.
\label{Classification Framework}
An overview of MESc can be seen in Figure \ref{fig:proposed_model}. The stages as detailed below:

An input document $D$ is tokenized into a sequence of tokens, $D = \{t_{1,D},t_{2,D},\cdots,t_{L_D,D}\}$ via a tokenizer specific to a chosen pre-trained language model (BERT, GPT etc.), where $t \in \mathbb{N}$ and $\mathbb{N}$ is the vocabulary of the tokenizer. 
This token sequence is split into a set of blocks $\{C_{1,D},C_{2,D},\cdots,C_{N_D,D}\}$ with overlaps($o$) with the previous block, which we call as chunks. Where each chunk block, $C_{i,D} = \{t_{(i+c-o),D},\cdots,t_{(i+2c-o,D}\}$ with $c$
 being the maximum number of tokens in the chunks, which is a predefined parameter for MESc (e.g. 512). $N_D = \lceil \frac{L_D}{c-o} \rceil$ is the total number of chunks for a document having $L$ tokens in total, with $o << c$. $N_D$ varies with the length of the document. 
 
    \paragraph{\textbf{Stage 1 - Custom fine-tuning:}} 
    \label{stage1}
    To each chunk of a document, we associate the document label $l_D$ and combine them together to form a token matrix:
    \begin{equation} I_D \in \mathbb{R}^{N_D \times c \times 1} \leftarrow [\{C_{1,D},l_D\},\{C_{2,D},l_D\},\cdots,\{C_{N_D,D},l_D\}] \end{equation}
    This is used as input for the document for fine-tuning the pre-trained encoder, where $N_D$ is the batch size for one pass through the encoder. 
        
    This allows the encoder to adapt to the domain-specific legal texts, which helps get richer features for the next stage.
    
    \paragraph{\textbf{Stage 2 - Extracting chunk embeddings:}}
    \label{level(2)}
    For a document, we pass its chunks $C_i$ through the fine-tuned encoder and extract its representation embeddings ($E_{i,D}$) from the last $l$ layers. $E_{i,D} \in \mathbb{R}^{l \times d}$, where $d$ is the dimension of the feature-length (we use $l=4$).
    The representation embeddings can be either the first token (as in BERT) or the last token for causal language models (as in GPT). We accumulate all $E_{i,D}$ of a document to form an embedding matrix:
    \begin{equation} 
    E_D \in \mathbb{R}^{N_D \times l \times d} \leftarrow \bigl[E_{1,D},E_{2,D},\cdots,E_{N_D,D}\bigr] 
    \end{equation}
    The $E_{i,D}$ acts as a representation of the chunk in this context, and combining them yields an approximate representation of the entire document.
    Doing this for all the documents gives us generated training data.

    \paragraph{\textbf{Stage 3 - Processing the extracted representations:}} \label{level(3)}
    Since the features extracted from the last layers of a fine-tuned encoder have different embedding spaces, they can contribute to being either positive or redundant. So for this stage, we choose to combine together the last $p<l$ layers in $E_D$ for further training. 
    We experiment with different $p$ before fixing one value as shown
    This gives $E_D^{(p)} \in \mathbb{R}^{N_D \times p \times d}, p \in \{1,2,3,4\}$.
    (We used 1,2 and 4 in our experiments to compare their effects.)
    We concatenate together the representations from these $p$ layers to get,
    \begin{equation} 
    E_{i,D}^{(p)} \in \mathbb{R}^{pd \times 1} \leftarrow \bigl[E_{i,D}^{(l)}|E_{i,D}^{(l-1)}|\cdots|E_{i,D}^{(l-p)}\bigr] 
    \end{equation}
    This gives,
    \begin{equation} 
    \widehat{E}_{D}^{(p)} \in \mathbb{R}^{N_D \times pd} \leftarrow \bigl\{E_{1,D}^{(p)}|E_{2,D}^{(p)}|\cdots|E_{N,D}^{(p)}\bigr\}\ 
    \end{equation}
    We also experimented with the element-wise addition of representations in $E_D^{(p)}$ and found their performance to be lower by 1 point in most of the experiments of section \ref{results}, hence we exclude it in MESc.
    \begin{enumerate}
        \item \textbf{Approximating the structure labels ($S_D$)} (Unsupervised learning): \label{stage_clustering}
        To get the information on the document's structure i.e. its parts (facts, arguments, concerned laws, etc.), we use a clustering mechanism (HDBSCAN \cite{hdbscan}). We cluster the $p$ chosen extracted chunk embeddings, $\widehat{E}_{D}^{(p)}$ to map similar parts of different documents together where the labels of one part of a document are learned by its similarity with another part of another document. The idea is that the embeddings of similar parts from different documents will group together forming a pool of cluster labels that can help identify its part in the document. One such dummy example can be seen in figure \ref{fig:cls_clustering}, where the $E_{i,D}^{(p)}$ of documents 1 and 2 learn their cluster (label) pool for, arguments of one type, 
        $a_1$ = \{$E^{(p)}_{1,1}, E^{(p)}_{1,2} E^{(p)}_{2,1}$\},
        facts of one type, 
        $f_1$ = \{$E^{(p)}_{1,3},E^{(p)}_{1,5},E^{(p)}_{2,2},E^{(p)}_{2,3},E^{(p)}_{2,5}$\},
        facts of another type, 
        $f_2$ =\{$E^{(p)}_{1,4},E^{(p)}_{1,6},E^{(p)}_{2,4},E^{(p)}_{2,6}$\}.
        So for document 1 the approximated structure then becomes $S_1$ = \{$a_1$,$a_1$,$f_1$,$f_2$,$f_1$,$f_2$\} and for document 2 it is $S_2$ = \{$a_1$,$f_1$,$f_1$,$f_2$,$f_1$,$f_2$\}. It is to be noted that this distinction if it's a fact or an argument etc. is done here for the purpose of representation. In the actual setting, it is unknown and these structure labels don't carry any specific name or meaning except for the model to give an understanding of its structure. 
        \begin{figure}[h]
          \centering
          \includegraphics[height=6.5cm, width=5cm]{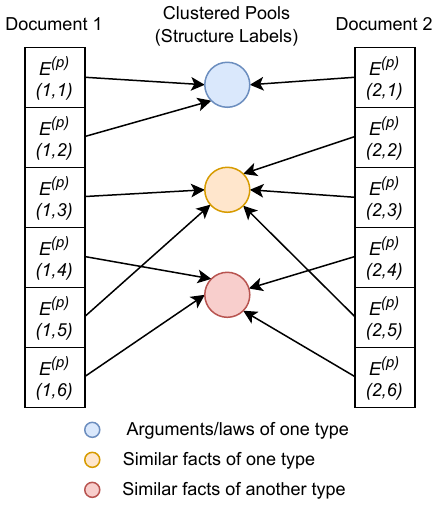}
          \caption{An example of clustering of chunk representations of two documents to generate structure labels.}
          \label{fig:cls_clustering}
        \end{figure}
        
        Since the performance of the HDBSCAN clustering mechanism decreases significantly with an increase in data dimension, we use a dimensionality reduction algorithm (pUMAP \cite{umap}), before clustering. For all the chunks of a document, their approximated structure labels are combined with the output of stage 3 (2), before processing through the final classification stage (4).

        \item \textbf{Global document representation} (Supervised learning):
        For intra-chunk attention, we use transformer encoder layers (Vaswani et al. \cite{transformer}), for one chunk to attend to another through its multi-head attention with a feed-forward neural network (FFN) layer. With respect to a chunk's position in the document, we add its positional embeddings (\cite{bert}) in ${E}_{D}^{(p)}$ and process it through $e$ transformer layers $T^{(e)}_{\{h, d_f\}}$, with $h$ attention heads and $d_f = pd$ as the dimension of the FFN. $e$ and $h$ are both hyperparameters whose choice depends upon the input feature lengths (Section \ref{results} evaluates different values of these parameters). 
    But $e\geq3$ sometimes overfits the model in our experiments, hence we fix  $e=2$ for MESc. The output is max-pooled and passed through a feed-forward neural network $FFN_T$ of $128$ nodes to get:
    \begin{equation} 
    G\left(\widehat{E}_{D}^{(p)}\right) = FFN_T\left(maxpool\left(T^{(e)}_{\{h, d_f\}}\left(\widehat{E}_{D}^{(p)}\right)\right)\right) \in \mathbb{R}^{128}
    \end{equation}

    \end{enumerate}
    \paragraph{\textbf{Stage 4 - Classification:}} \label{Level4}The structure labels along with the output of the feed-forward network of stage 3(b) are concatenated together. Which is processed through an internal feed-forward network $FFN_i$ (32 nodes, with softmax activation) and  an external feed-forward network $FFN_e$ ($u$ label(class) nodes with task-specific activation function sigmoid(softmax)) to get the output $O(D)$ for a document $D$, as shown in equation \ref{eq.O}. 
    \begin{equation} \label{eq.O}
    O\left(D\right) = FFN_e\left( FFN_i \left( \left(\left[G\left(\widehat{E}_{D}^{(p)}\right)|S_D\right]\right) \right) \right) \in \mathbb{R}^{u}
    \end{equation}
    $O$ and $G$ are learnt together for final classification while $S_D$ is learnt independently.

\subsection{Extractive Explanation Algorithm - "Occlusion-based Relevant Sentence Extraction (ORSE)"} 
To understand the relevant parts of the document contributing to the said decision prediction, we extract sentences that have a high impact on the decision prediction. We develop an 
extractive explanation algorithm for hierarchical classification models (similar to MESc, \cite{Hierarchical_Transformers}) based on the input sensitivity of the models in the hierarchy. Since this is an extractive process there is no need for a \say{pre-summarised} (detailed annotation) explanation, which is required to train an explanation model.
A hierarchical model is composed of many individual models divided into levels, where each level is responsible to learn different components of the input, which are combined together in moving up the hierarchy. 
Taking an example of a document, the input in these models can be processed in a hierarchical fashion from the bottom up by learning from the words, then combining them into sentences, and into paragraphs/parts which can be further accumulated to give a full input representation.

In this work, we target ORSE to explain the reason why a decision prediction is made by our hierarchical predictive model (MESc) where the algorithm processes the document in its two steps of hierarchy. 

1) Find the highly sensitive (impactful) chunks (parts of the document).

2) Find the highly sensitive sentences from these chunks.

We define ORSE (Algorithm \ref{alg:Explanation-algo}) keeping in mind the long lengths of the large documents. While it can also be adapted to shorter documents (a few hundred to thousand tokens), for which the extraction can be done by just the 1st step.


For a classification model $M$ and input $I={\{i_j|0 \leq j \leq n}\}$, where $n$ is the input length, consider $M(I) = O_I$ to be the prediction without any occlusion where $P$ is the final predicted class label(s), and $M(\{I|i_j\})=O_I^{(j)}$ to be the prediction after the occlusion of $i_j$ in $I$. The occlusion is done by masking individual parts (e.g. $0$ masking) before feeding into a classification model. If $P$ is from the final model in the hierarchy we take it as the absolute class label with which we rank the inputs for all models in the hierarchy. 
We define an occlusion-sensitivity impact function $L$, depending on the classification problem type as,
\begin{equation}\label{impact_score}
    L(O_I^{(j)},P) = 
    \begin{cases}
    CCE_{loss}(O_I^{(j)},P) ,& \text{for multi-class} \\
    BCE_{loss}(O_I^{(j)},P) ,& \text{for binary, multi-label}
    \end{cases}
\end{equation}
where, $CCE_{loss}=$ Categorical-Cross-Entropy loss
, and $BCE_{loss}=$ 
Binary-Cross-Entropy loss. Other loss functions can also be used depending on the task. The intuition behind this impact measure is to see how important is the occluded part of the input for a prediction with the change in its loss function. A higher loss means more impact. $L$ is to be chosen such that it is always $>0$.

To rank these losses in terms of their impact, we measure the deviance of an input's occluded component's impact value from the impact value of the whole input (without any occlusion).
With $L$ we compute the \say{weighted occlusion sensitivity score} $S$, 
\begin{equation}\label{occ_score}
    S(s,L_j,L_I) = s \times \left(L_I^{(j)} - L_I + \delta \right)  
\end{equation}
where $s$ is the score weight, $L_I^{(j)}$ and $L_I$ are impact scores on occluding $i_j$ in $I$ and impact of $I$ with the $P$ (absolute class label). We add constant $\delta$ so as to make $S$ positive by shifting the axis, which is required to keep the score above $0$. 

\begin{algorithm}\footnotesize
\caption{Occlusion sensitivity-based Relevant Sentence Extractor (ORSE)}\label{alg:Explanation-algo}
\begin{algorithmic}[1]
\Require From \ref{Classification Framework}, Select a classification model $M$, and its backbone fine-tuned encoder $T$. $k = \%$ of sentences to choose.
\For{all documents}
\State Divide the document into chunks of length $c$.
    \State $E \gets$ Extract all chunk embeddings from $T$. 
    \State $O_E \gets M(E)$, probability output.
    \State $P \gets$ absolute predicted label from $O_E$
    \State $L_E \gets 0$, impact with itself $L(P,P)$ (Eq. \ref{impact_score}) 
    \For{chunk $c_i$ in $E$}
    \State Mask $c_i$
    \State $O_E^{(i)} \gets M(\{E|c_i\})$, probability output after masking $c_i$.
    \State $L_{c_i} \gets L(O_E^{(i)},P)$ (Eq. \ref{impact_score})
    \State $S_{c_i} \gets S(1,L_{c_i},L_E)$ (Eq. \ref{occ_score})
    \EndFor
    \State $S_E \gets$ concatenate all $(c_i, S_{c_i})$.
    \State $S_E \gets$ Sort $S_E$ in descending order of $S_{c_i}$.
    \For{($C_i$, $s$) in $S_E$}
        \State $O_{C_i} \gets T(C_i)$, probability output from $T$
        \State $L_{C_i} \gets L(O_{C_i},P)$ (Eq. \ref{impact_score})
        \State Split $C_i$ into sentences, $\{s_j|1\leq j\leq$ total sentences$\}$.
        \For{$s_j$ in $C_i$}
        \State Mask $s_j$.
        \State $O_{s_j} \gets T(\{C_i|s_j\})$, probability after masking $s_j$.
        \State $L_{s_j} \gets L(O_{s_j},P)$ (Eq. \ref{impact_score})
        \State $S_{s_j} \gets S(s,L_{s_j},L_{C_i})$ (Eq. \ref{occ_score})
        \State $A_{score}$ $\gets$ concatenate all $(i,s_j,S_{s_j})$.
        \EndFor
    \EndFor
    \State Sort $A_{score}$ in descending order of $S_{s_j}$. 
    \State $A_{score}[k] \gets$ keep the top $k \%$ sentences.
    \State $A_{score}[k] \gets$ rearrange in the order of (i,$s_j$).
\EndFor
\end{algorithmic}
\end{algorithm}

  We give a description of ORSE adapted to MESc in the Algorithm \ref{alg:Explanation-algo}, and detail the steps involved. 
  We start from the top-level model $M$ (stage 4) to find the highly sensitive chunks (steps 2-14), for a document. We calculate the probability output from $M$. Since $M$ is at the top level of the hierarchy we take its prediction as the absolute predicted label $P$, and take the self-impact score as $0$ (Step 2-6). We mask/occlude the chunks and calculate their impact score (Eq. \ref{impact_score}) and then their weighted occluded sensitivity scores (Eq. \ref{occ_score}) with respect to the whole document i.e. self-impact score (steps 8-11). Since this is the top level we use $1$ as the weight. We sort the accumulated scores in order of their sensitivity score (i.e. higher value is given more importance). 

 To rank the sentences (steps 15-28) we iteratively start from the highest-scored chunk and take its probability output from the fine-tuned transformer $T$ (from stage 1) to calculate its impact-score w.r.t $P$ (step 17). 
 We then split this chunk ($c_i$) into sentences and iteratively mask/occlude a sentence $s_j$ inside the chunk to calculate its weighted occluded sensitivity score ($S_{s_j}$) (steps 19-24). To weigh the overall importance of each sentence of this chunk as compared to the sentences belonging to other chunks, we weigh the impact shift of $s_i$ with the sensitivity score of $C_i$ from the previous level of hierarchy. We store the sentences along with their chunk number and sensitivity score in $A_{score}$. 
 We sort $A_{score}$, ranking in the order of $S_{s_j}$. Since this is the last level of the hierarchy we stop and take the top $k\%$ sentences. To arrange the sentences with their sequential occurrence in the document we arrange $A_{score}[k]$ according to the chunk number and the sentence in the chunk. These sentences serve as the explanation for a document's prediction.
 The time complexity is model dependent, and is $O(n^2)$ here, due to the quadratic complexity of the fine-tuned transformer ($T$) used, where asymptotically $n$ is the average length of all the documents for a batch. 

\section{Experimental setup}
\label{Experimental setup}
For our backbone transformer encoder, we used the domain-specific pre-trained model LEGAL-BERT\cite{legal-bert}, InLegalBERT\cite{InLegalBERT} and chose GPT-Neo\cite{gpt-neo}, GPT-J\cite{gpt-j} for experimenting with larger LLMs with multi-billion parameters. The tokenizers used to tokenize the documents are from the same encoders. No chunks were excluded in any stage of MESc.

\textbf{Stage 1:} For BERT-based encoders, the chunk size was set to $\leq510$ (90 token overlaps) with global tokens ([CLS],[SEP]) and padding to make the input chunk size $512$. 
For GPT-based encoders, we experimented with the same overlap but two different chunk sizes $512$ and $2048$, the latter being the maximum possible input length. 
We abbreviate the encoders fine-tuned on 512 input length as ($\alpha$) and for ones fine-tuned with 2048 input length as ($\gamma$). 

For testing the fine-tuned models, we considered the last tokens for each document. For all GPT-Neo($\alpha$) and GPT-J($\alpha$) we evaluated on input length of 512 tokens to compare their performance with 
LEGAL-BERT($\alpha$) and InLegalBERT($\alpha$).
 
These encoders were fine-tuned for 4 epochs with learning rates in $\{2e^{-6}$,$3e^{-5}\}$, and we chose the best-performing one for stage 2. 
For the full-fine-tuning of the GPT-J and all GPT-Neo, we used 6 Nvidia A100 (80GB GPU) with ZeRO-3 optimization strategy implemented in Deepspeed\footnote{\url{https://www.deepspeed.ai/}} with huggingface's Accelerate library\footnote{\url{https://huggingface.co/docs/accelerate/index}}.

\textbf{Stage 2:} We used the [CLS] token for embeddings extraction in BERT-based models and for the causal language models(GPT-Neo and GPT-J) we used the last token's embedding as the global representation for the respective chunk. 

\textbf{Stage 3 \& 4:} Adam optimizer (learning rate = $3.5e^{-6}$) was used after stage 2 of the MESc. For the binary and multi-label classification problems, we used \say{binary cross-entropy loss} and used \say{categorical cross-entropy loss} respectively. We experimented with $N=\{1,2,3\}$ transformer encoder layers in stage 3 
with $h = 8$ and trained for 5 epochs, and chose $N=2$ after analyzing their performance (section \ref{results}). For clustering to approximate our structure labels we use HDBSCAN (minimum cluster size = $15$) and pUMAP\footnote{\url{https://umap-learn.readthedocs.io/en/latest/parametric_umap.html}}($64$ output dimension).
,

 The best-performing MESc configuration for ILDC (Table \ref{tab:results_MESc_ildc}), with $512$ input chunks and $k=\{0.2,0.3,0.4\}$ was used for ORSE (Table \ref{tab:Explanation_res}). 


\subsection{Dataset}
\label{dataset_desc}
We choose the legal datasets having large documents with a nonuniform structure throughout, without any structural annotations or information. 
Suiting our problem of large scarce-annotation documents we found one such dataset in the Indian legal court setting, named IDLC \cite{ildc-cjpe}, and the same requirement in a subset of the LexGLUE dataset \cite{lexglue}. The ILDC dataset includes highly unstructured 39898 English-language case transcripts from the Supreme Court of India (SCI), where the final decisions have been removed from the document from the end. Upon analyzing the documents from their sources and the dataset we found that they are highly unstructured and noisy. 
The initial decision between "rejected" and "accepted" made by the SCI judge(s) is used to identify each document that also serves as their decision label. To assess how well the judgment prediction algorithms explain themselves, a piece of the corpus (ILDC$_{Expert}$, a separate test set of 56 documents) is labeled with gold standard explanations by five distinct legal experts which are pieces of text selected from the document that is most relevant to the judgment. We use this to evaluate our extractive explanation algorithm ORSE.

The LexGLUE dataset \cite{lexglue} comprises a set of seven datasets from the European Union and US court case setting, for uniformly assessing model performance across a range of legal NLP tasks, from which we choose ECtHR (Task A), ECtHR (Task B), and SCOTUS as they are classification tasks involving our problem of long unstructured legal documents. ECtHR (A and B) are court cases from European Convention on Human Rights (ECHR) for articles that were violated or allegedly violated. The dataset contains factual paragraphs from the description of the cases. SCOTUS consists of court cases from the highest federal court in the United States of
America, with metadata from SCDB \footnote{\url{http://scdb.wustl.edu/}}. The details of the number of labels and the average and maximum document length (in tokens) with task description can be found in the table \ref{dataset_des}. The tokenization in table \ref{dataset_des} is done using the tokenizer of GPT-J. 

The ILDC, ECtHR (Task A), ECtHR (Task B), and SCOTUS serve as a good fit for our problem and test our approach.
For performance comparison on LexGLUE, we used the SOTA benchmark of Chalkidis et al. \cite{lexglue}, and for ILDC we used the benchmark from its paper \cite{ildc-cjpe} and of Shounak et al. \cite{InLegalBERT}'s experiments on their models specifically pre-trained on the Indian legal cases.
Since LexGLUE lacks an explanation set as like in ILDC$_{expert}$ we couldn't use ORSE to test its effectiveness on LexGLUE.

\begin{table}
\centering
\caption{Dataset statistics}
\begin{adjustbox}{width=\columnwidth}
\label{dataset_des}
\begin{tabular}{|c|ccc|ccc|c|c|} \hline
\multirow{2}{*}{Name} & \multicolumn{3}{c|}{No. of Documents} & \multicolumn{3}{c|}{\begin{tabular}[c]{@{}c@{}}Average length \\\&\\Maximum length\\(tokens)\end{tabular}} & \multirow{2}{*}{\begin{tabular}[c]{@{}c@{}}No. of \\labels\end{tabular}} & \multirow{2}{*}{\begin{tabular}[c]{@{}c@{}}Problem\\Type\end{tabular}} \\ \cline{2-7}
 & Train & Validation & Test & Train & Validation & Test &  &  \\ \hline
ILDC & 37387 & 994 & 1517 & \begin{tabular}[c]{@{}c@{}}4120\\501275\end{tabular} & \begin{tabular}[c]{@{}c@{}}5104\\58048\end{tabular} & \begin{tabular}[c]{@{}c@{}}5238\\55703\end{tabular} & 2 & Binary \\ \hline
ECtHR(A) & 9000 & 1000 & 1000 & \begin{tabular}[c]{@{}c@{}}2011\\46500\end{tabular} & \begin{tabular}[c]{@{}c@{}}2210\\18352\end{tabular} & \begin{tabular}[c]{@{}c@{}}2401\\20835\end{tabular} & 10 & Multi-Label \\ \hline
ECtHR(B) & 9000 & 1000 & 1000 & \begin{tabular}[c]{@{}c@{}}2011\\46500\end{tabular} & \begin{tabular}[c]{@{}c@{}}2210\\18352\end{tabular} & \begin{tabular}[c]{@{}c@{}}2401\\20835\end{tabular} & 10 & Multi-Label \\ \hline
SCOTUS & 5000 & 1400 & 1400 & \begin{tabular}[c]{@{}c@{}}8291\\126377\end{tabular} & \begin{tabular}[c]{@{}c@{}}12639\\56310\end{tabular} & \begin{tabular}[c]{@{}c@{}}12597\\124955\end{tabular} & 13 & Multi-Class \\ \hline
\end{tabular}
\end{adjustbox}
\end{table}

\section{Results and discussion}
\label{results}
\begin{table*}[h] 
\caption{Custom-finetuning results on the chosen pre-trained transformer encoder language models (in Section \ref{Experimental setup}), e = epoch}
\begin{adjustbox}{width={\textwidth}}\footnotesize
\label{tab:finetuning_result}
\begin{tabular}{|c|c|c|c|c|c|c|c|c|} \hline
\multicolumn{9}{|l|}{$\alpha$: fine-tuned and evaluated with 512 input length, $\beta$: evaluating $\alpha$ on its maximum input length, $\gamma$: fine-tuned and evaluated with maximum input length.} \\ \hline
\multirow{2}{*}{Dataset} & \multicolumn{2}{c|}{\begin{tabular}[c]{@{}c@{}}LEGAL-BERT\\($\mu$-F1/m-F1)\end{tabular}} & \multicolumn{2}{c|}{\begin{tabular}[c]{@{}c@{}}GPT-Neo 1.3B\\($\mu$-F1/m-F1)\end{tabular}} & \multicolumn{2}{c|}{\begin{tabular}[c]{@{}c@{}}GPT-Neo 2.7B\\($\mu$-F1/m-F1)\end{tabular}} & \multicolumn{2}{c|}{\begin{tabular}[c]{@{}c@{}}GPT-J 6B\\($\mu$-F1/m-F1)\end{tabular}} \\ \cline{2-9}
 & Validation & Test & Validation & Test & Validation & Test & Validation & Test \\ \hline
ECtHR (A) & \begin{tabular}[c]{@{}c@{}}($\alpha$) 0.6408/0.5095\\(e = 4)\end{tabular} & \begin{tabular}[c]{@{}c@{}}($\alpha$) 0.6285/0.4866\\(e = 4)\end{tabular} & \begin{tabular}[c]{@{}c@{}}($\alpha$) 0.6705/0.5940\\($\beta$) 0.6708/0.5958\\(e = 2)\end{tabular} & \begin{tabular}[c]{@{}c@{}}($\alpha$) 0.6619/0.5659\\($\beta$) 0.6620/0.5716\\(e = 2)\end{tabular} & \begin{tabular}[c]{@{}c@{}}($\alpha$) 0.6815/0.5833\\($\beta$) 0.6739/0.5947\\(e = 2)\end{tabular} & \begin{tabular}[c]{@{}c@{}}($\alpha$) 0.6849/0.5445\\($\beta$) 0.6811/0.5649\\(e = 2)\end{tabular} & \begin{tabular}[c]{@{}c@{}}($\alpha$) 0.7260/0.6715\\($\beta$) 0.7567/0.6945\\($\gamma$) \textbf{0.7855/0.7550}\\(e = 3)\end{tabular} & \begin{tabular}[c]{@{}c@{}}($\alpha$) 0.7142/0.5927\\($\beta$) 0.7330/0.6245\\($\gamma$) \textbf{0.7451/0.6467}\\(e = 3)\end{tabular} \\ \hline
ECtHR (B) & \begin{tabular}[c]{@{}c@{}}($\alpha$) 0.6961/0.6255\\(e = 3)\end{tabular} & \begin{tabular}[c]{@{}c@{}}($\alpha$) 0.7089/0.6405\\(e = 3)\end{tabular} & \begin{tabular}[c]{@{}c@{}}($\alpha$) 0.7459/0.6938\\($\beta$) 0.7542/0.6946\\(e = 2)\end{tabular} & \begin{tabular}[c]{@{}c@{}}($\alpha$) 0.7542/0.7091\\($\beta$) 0.7574/0.7009\\(e = 2)\end{tabular} & \begin{tabular}[c]{@{}c@{}}($\alpha$) 0.7524/0.7147\\($\beta$) 0.7619/0.7252\\(e = 2)\end{tabular} & \begin{tabular}[c]{@{}c@{}}($\alpha$) 0.7448/0.6826\\($\beta$) 0.7513/0.7072\\(e = 2)\end{tabular} & \begin{tabular}[c]{@{}c@{}}($\alpha$) 0.7769/0.7244\\($\beta$) 0.8069/0.7611\\($\gamma$) \textbf{0.8308/0.8039}\\(e = 3)\end{tabular} & \begin{tabular}[c]{@{}c@{}}($\alpha$) 0.7715/0.7326\\($\beta$) 0.8049/0.7631\\($\gamma$) \textbf{0.8316/0.7927}\\(e = 3)\end{tabular} \\ \hline
SCOTUS & \begin{tabular}[c]{@{}c@{}}($\alpha$) 0.7296/0.5924\\(e = 6)\end{tabular} & \begin{tabular}[c]{@{}c@{}}($\alpha$) 0.6876/0.5357\\(e = 6)\end{tabular} & \begin{tabular}[c]{@{}c@{}}($\alpha$) 0.7300/0.6582\\($\beta$) 0.7614/0.6772\\($\gamma$) 0.7731/0.6830\\(e = 2)\end{tabular} & \begin{tabular}[c]{@{}c@{}}($\alpha$) 0.7114/0.6035\\($\beta$) 0.7371/0.6310\\($\gamma$) 0.7502/0.6438\\(e = 2)\end{tabular} & \begin{tabular}[c]{@{}c@{}}($\alpha$) 0.7314/0.6571\\($\beta$) 0.7686/0.6851\\($\gamma$) 0.7828/0.6931\\(e = 1)\end{tabular} & \begin{tabular}[c]{@{}c@{}}($\alpha$) 0.7057/0.6025\\($\beta$) 0.7364/0.6564\\($\gamma$) 0.7636/0.6619\\(e = 1)\end{tabular} & \begin{tabular}[c]{@{}c@{}}($\alpha$) 0.7592/0.6875\\($\beta$) 0.7950/0.7295\\($\gamma$) \textbf{0.8178/0.7513}\\(e = 3)\end{tabular} & \begin{tabular}[c]{@{}c@{}}($\alpha$) 0.7200/0.6276\\($\beta$) 0.7571/0.6625\\($\gamma$) \textbf{0.7850/0.7196}\\(e = 3)\end{tabular} \\ \hline
\multicolumn{1}{|l|}{} & \multicolumn{2}{c|}{InLegalBERT (accuracy(\%)/m-F1)} & \multicolumn{6}{c|}{accuracy(\%)/m-F1} \\ \hline
ILDC & \begin{tabular}[c]{@{}c@{}}($\alpha$) \textbf{76.15/76.8}\\(e = 4)\end{tabular} & \begin{tabular}[c]{@{}c@{}}($\alpha$) \textbf{76.00/76.10}\\(e = 4)\end{tabular} & \begin{tabular}[c]{@{}c@{}}($\alpha$) 74.25/0.7421\\($\beta$) 76.66/0.7662\\(e=1)\end{tabular} & \begin{tabular}[c]{@{}c@{}}($\alpha$) 72.91/0.7291\\($\beta$) 77.26/0.7725\\(e=1)\end{tabular} & \begin{tabular}[c]{@{}c@{}}($\alpha$) 76.96/0.7675\\($\beta$) 81.59/0.8144\\(e=1)\end{tabular} & \begin{tabular}[c]{@{}c@{}}($\alpha$) 74.29/0.7424\\($\beta$) 81.21/0.8118\\(e=1)\end{tabular} & \begin{tabular}[c]{@{}c@{}}($\alpha$) 75.15/0.7511\\($\beta$) 79.78/0.7972\\($\gamma$)~\textbf{83.60/0.8347}\\(e=1)\end{tabular} & \begin{tabular}[c]{@{}c@{}}($\alpha$) 73.96/0.7396\\($\beta$) 81.93/0.8192\\($\gamma$) \textbf{83.72/0.8366}\\(e=1)\end{tabular} \\ \hline
\end{tabular}
\end{adjustbox}
\end{table*}

\begin{table*}[h] \footnotesize
\centering
\caption{Results on LexGLUE for different configurations of MESc (* is the encoder model used for embedding extraction)}
\label{results_table_lexglue}
\begin{tabular}{|l|c|c|c|c|c|c|c|} \cline{3-8}
\multicolumn{2}{l|}{} & \multicolumn{2}{c|}{ECtHR (A)} & \multicolumn{2}{c|}{ECtHR (B)} & \multicolumn{2}{c|}{SCOTUS} \\ \hline
\multicolumn{1}{|c|}{\multirow{2}{*}{\begin{tabular}[c]{@{}c@{}}$p$ layers,\\$e$ x Encoder\end{tabular}}} & \multirow{2}{*}{Structure Labels} & Validation & Test & Validation & Test & Validation & Test \\ \cline{3-8}
\multicolumn{1}{|c|}{} &  & \multicolumn{6}{c|}{$\mu$-F1/m-F1} \\ \hline
\multicolumn{2}{l|}{LexGLUE benchmark \cite{lexglue}} & 0.725/0.682 & 0.712/0.647 & 0.797/0.768 & 0.804/0.747 & 0.776/0.633 & 0.766/0.665 \\ \hline
\multicolumn{2}{l|}{LEGAL-BERT* ($\alpha$)} & \multicolumn{1}{c}{} &  & \multicolumn{1}{c}{} &  & \multicolumn{1}{c}{} &  \\ \hline
$p$=1, 1 x & No & 0.7005/0.6118 & 0.6825/0.5806 & 0.7470/0.6791 & 0.7418/0.6890 & 0.7719/0.6926 & 0.7136/0.5916 \\ \hline
$p$=1, 2 x & No & 0.6984/0.6056 & 0.6923/0.5935 & 0.7507/0.6835 & 0.7386/0.6742 & 0.7729/0.6895 & 0.7152/0.5817 \\ \hline
\multirow{2}{*}{$p$=4, 1 x} & No & 0.7718/0.6994 & 0.7546/0.6226 & 0.8084/0.7709 & 0.8102/0.7573 & 0.7928/0.6866 & 0.7396/0.5865 \\ \cline{2-8}
 & Yes & 0.7652/0.6870 & 0.7582/0.6378 & 0.8087/0.7727 & 0.8122/0.7725 & 0.7959/0.7025 & 0.7525/0.6194 \\ \hline
\multirow{2}{*}{$p$=4, 2 x} & No & 0.7662/0.6479 & 0.7543/0.6337 & 0.8078/0.7574 & 0.8118/0.7564 & 0.7899/0.6849 & 0.7431/0.6054 \\ \cline{2-8}
 & Yes & 0.7682/0.6883 & \textbf{0.7618/0.6508} & \textbf{0.8089/0.7748} & \textbf{0.8157/0.7670} & \textbf{0.7952/0.6872} & \textbf{0.7550/0.6208} \\ \hline
\multirow{2}{*}{$p$=4, 3 x} & No & \textbf{0.7884/0.6905} & 0.7523/0.6311 & 0.8138/0.7863 & 0.8132/0.7699 & 0.7928/0.6866 & 0.7399/0.5635 \\ \cline{2-8}
 & Yes & 0.7714/0.6815 & 0.7510/0.6309 & 0.8075/0.7564 & 0.8100/0.7621 & 0.7729/0.6536 & 0.7392/0.5783 \\ \hline
\multicolumn{2}{l|}{Gpt-Neo 1.3B* ($\alpha$)} & \multicolumn{1}{c}{} &  & \multicolumn{1}{c}{} &  & \multicolumn{1}{c}{} &  \\ \hline
\multirow{2}{*}{$p$=2, 2 x} & No & 0.7231/0.6862 & 0.7115/0.6359 & 0.7960/0.7307 & 0.8030/0.7702 & 0.7862/0.7185 & 0.7536/0.6479 \\ \cline{2-8}
 & Yes & \textbf{0.7407/0.7033} & \textbf{0.7273/0.6448} & \textbf{0.7939/0.7479} & \textbf{0.8040/0.7808} & \textbf{0.7797/0.7284} & \textbf{0.7646/0.6592} \\ \hline
\multirow{2}{*}{$p$=4, 2 x} & No & 0.7358/0.7059 & 0.7146/0.6277 & 0.7947/0.7483 & 0.8086/0.7664 & 0.7722/0.7026 & 0.7429/0.6352 \\ \cline{2-8}
 & Yes & 0.7248/0.7076 & 0.7068/0.6410 & 0.7968/0.7537 & 0.8060/0.7757 & 0.7651/0.7036 & 0.7418/0.6377 \\ \hline
\multicolumn{2}{l|}{GPT-Neo 2.7B* ($\alpha$)} & \multicolumn{1}{c}{} &  & \multicolumn{1}{c}{} &  & \multicolumn{1}{c}{} &  \\ \hline
\multirow{2}{*}{$p$=2, 2 x} & No & 0.7380/0.6750 & 0.7457/0.6224 & 0.7896/0.7689 & 0.7949/0.7620 & 0.7845/0.7140 & 0.7676/0.6570 \\ \cline{2-8}
 & Yes & \textbf{0.7634/0.7105} & \textbf{0.7567/0.6644} & \textbf{0.7986/0.7693} & \textbf{0.8072/0.7696} & \textbf{0.7988/0.7348} & \textbf{0.7627/0.6630} \\ \hline
\multirow{2}{*}{$p$=4, 2 x} & No & 0.7510/0.6641 & 0.7524/0.6355 & 0.7897/0.7599 & 0.7940/0.7503 & 0.7818/0.7273 & 0.7577/0.6554 \\ \cline{2-8}
 & Yes & 0.7600/0.6857 & 0.7587/0.6561 & 0.7825/0.7686 & 0.7935/0.7635 & 0.7903/0.7245 & 0.7641/0.6775 \\ \hline
\multicolumn{2}{l|}{GPT-J 6B* ($\alpha$)} & \multicolumn{1}{c}{} &  & \multicolumn{1}{c}{} &  & \multicolumn{1}{c}{} &  \\ \hline
\multirow{2}{*}{$p$=2, 2 x} & No & 0.7516/0.7138 & 0.7222/0.6263 & 0.7997/0.7588 & 0.7931/0.7692 & 0.7888/0.7215 & 0.7505/0.6658 \\ \cline{2-8}
 & Yes & \textbf{0.7529/0.7255} & \textbf{0.7163/0.6406} & \textbf{0.8048/0.7722} & \textbf{0.7977/0.7760} & \textbf{0.7791/0.7343} & \textbf{0.7598/0.6715} \\ \hline
\multirow{2}{*}{$p$=4, 2 x} & No & 0.7351/0.6743 & 0.7156/0.6118 & 0.7707/0.7414 & 0.7800/0.7605 & 0.7967/0.7211 & 0.7490/0.6333 \\ \cline{2-8}
 & Yes & 0.7544/0.7377 & 0.7219/0.6437 & 0.7891/0.7534 & 0.7795/0.7625 & 0.7872/0.7268 & 0.7485/0.6593 \\ \hline
\multicolumn{2}{l|}{GPT-J 6B* ($\gamma$)} & \multicolumn{1}{c}{} &  & \multicolumn{1}{c}{} &  & \multicolumn{1}{c}{} &  \\ \hline
\multirow{2}{*}{$p$=2, 2 x} & No & 0.7565/0.7182 & 0.7384/0.6434 & 0.8055/0.7954 & 0.8094/0.7675 & 0.8141/0.7508 & 0.7688/0.6773 \\ \cline{2-8}
 & Yes & \textbf{0.7619/0.7383} & \textbf{0.7470/0.6571} & \textbf{0.8201/0.8076} & \textbf{0.8169/0.7801} & \textbf{0.8164/0.7755} & 0.7814/0.6853 \\ \hline
\multirow{2}{*}{$p$=4, 2 x} & No & 0.7512/0.7007 & 0.7296/0.6333 & 0.8142/0.7937 & 0.8113/0.7763 & 0.8170/0.7533 & 0.7728/0.6786 \\ \cline{2-8}
 & Yes & 0.75860.7174 & 0.7484/0.6548 & 0.8192/0.7977 & 0.8134/0.7802 & 0.8248/0.7555 & \textbf{0.7867/0.6966} \\ \hline
\end{tabular}
\end{table*}

\begin{table}[h] \footnotesize
\caption{Results on ILDC\cite{ildc-cjpe} for the best configurations of MESc (* is the encoder model used for embedding extraction)}
\label{tab:results_MESc_ildc}
\begin{tabular}{|c|c|c|c|} \cline{3-4}
\multicolumn{1}{l}{} & \multicolumn{1}{l|}{} & \multicolumn{2}{c|}{ILDC} \\ \cline{3-4}
\multicolumn{1}{c}{} & \multicolumn{1}{l|}{} & Validation & Test \\ \cline{3-4}
\multicolumn{1}{l}{} & \multicolumn{1}{l|}{} & \multicolumn{2}{c|}{Accuracy(\%)/m-F1} \\ \hline
\multicolumn{2}{l|}{Shounak et al. \cite{InLegalBERT} benchmark} & - & -/83.09 \\ \hline
\multicolumn{2}{l|}{ILDC \cite{ildc-cjpe} benchmark} & - & 77.78/77.79 \\ \hline
$p$ layers, $e$ x Encoder & \begin{tabular}[c]{@{}c@{}}Structure\\labels\end{tabular} & \multicolumn{1}{l|}{} & \multicolumn{1}{l|}{} \\ \hline
\multicolumn{2}{l|}{InLegalBERT* ($\alpha$)} & \multicolumn{1}{l|}{} & \multicolumn{1}{l|}{} \\ \hline
\multirow{2}{*}{$p$=1, 1 x} & No & 84.10/84.21 & 83.72/83.73 \\ \cline{2-4}
 & Yes & 84.51/84.53 & 83.65/83.65 \\ \hline
\multirow{2}{*}{$p$=1, 2 x} & No & 83.90/84.00 & 83.45/83.47 \\ \cline{2-4}
 & Yes & 85.11/85.15 & 83.78/83.78 \\ \hline
\multirow{2}{*}{$p$=4, 1 x} & No & 84.30/84.32 & 83.41/83.41 \\ \cline{2-4}
 & Yes & \textbf{85.23/85.25} & \textbf{84.15/84.15} \\ \hline
\multirow{2}{*}{$p$=4, 2 x} & No & 84.30/84.32 & 83.72/83.68 \\ \cline{2-4}
 & Yes & \textbf{85.15/85.17} & \textbf{84.11/84.13} \\ \hline
\multicolumn{1}{l|}{GPT-Neo 2.7B* ($\alpha$)} & \multicolumn{1}{l|}{} & \multicolumn{1}{l|}{} & \multicolumn{1}{l|}{} \\ \hline
\multirow{2}{*}{$p$=2, 2 x} & No & 84.13/84.12 & 82.97/82.79 \\ \cline{2-4}
 & Yes & 84.71/84.67 & 83.65/83.64 \\ \hline
\multirow{2}{*}{$p$=4, 2 x} & No & 84.10/84.09 & 83.01/83.00 \\ \cline{2-4}
 & Yes & 84.30/84.29 & 83.22/83.21 \\ \hline
\multicolumn{1}{l|}{GPT-J 6B* ($\alpha$)} & \multicolumn{1}{l|}{} & \multicolumn{1}{l|}{} & \multicolumn{1}{l|}{} \\ \hline
\multirow{2}{*}{$p$=2, 2 x} & No & 83.43/83.42 & 82.84/82.78 \\ \cline{2-4}
 & Yes & 84.32/84.31 & 83.21/83.19 \\ \hline
\multirow{2}{*}{$p$=4, 2 x} & No & 83.45/83.46 & 82.73/82.73 \\ \cline{2-4}
 & Yes & 84.22/84.21 & 83.37/83.36 \\ \hline
\end{tabular}
\end{table}

\subsection{Results on classification framework} 
$\mu$-F$1$ (micro) and $m$-F$1$ (macro) are used to measure the performance for the LexGLUE dataset. And accuracy(\%) and macro-F$1$ for the ILDC dataset. We emphasize more on the $\mu$-F$1$ for the LexGULE dataset taking into the class imbalance whilst we also take $m$-F$1$ into consideration to compare performance with previous benchmarks of LexGLUE \cite{lexglue}.
We list out the detailed experimental results for best configurations of MESc in table \ref{results_table_lexglue} and 
\ref{tab:results_MESc_ildc}, and the fine-tuned performance of the LLMs used in table \ref{tab:finetuning_result}. 
\subsubsection{\textbf{Intra-domain(legal) transfer learning:}} 
As can be seen from table \ref{tab:finetuning_result}, for LexGLUE's subset, all the GPTs used here are able to adapt better with a minimum of $\approx$ 3 points gain on $\mu$-F1 and a minimum of $\approx$ 6 points on m-F1 score.
On the other hand in the ILDC dataset, for the 512-fine-tuned variants with 512 input lengths for evaluation, their performance dropped or remained similar to the InLegalBERT, while upon increasing the evaluation input length to 2048 we can see an increase of more than 1 point in the performance. For GPT-J when fine-tuned with 2048 input length, the performance increase, compared to its 512 variant, is at least a minimum of $\approx$ 2 points for all the datasets. We can see that an increase in the input length for fine-tuning helps to capture more feature information for such documents. Also going from GPT-Neo-1.3B's 1.3 billion parameters to its 2.7 billion to 6 billion GPT-J the performance increases by a margin 
of 2 points at minimum, where we can see the parameter count playing an important role in adaptation and understanding these documents. 
Even though GPT-Neo and GPT-J are pre-trained on US legal cases (Pile \cite{pile}) they are able to adapt better to the European and Indian legal documents, with with a minimum gain of $\approx$ 7 points ($\gamma$) and the ECtHR(A \& B) and the ILDC dataset over their domain-specific pre-trained counterparts LEGAL-BERT and InLegalBERT respectively.
\subsubsection{\textbf{Performance with MESc:}}
\label{Performance with MESc}
Looking at table \ref{results_table_lexglue} and table \ref{tab:results_MESc_ildc} we interpret the results in two directions.
    \paragraph{\textbf{Encoders fine-tuned on 512 input length $(\alpha)$:}} 
    For LEGAL-BERT and InLegalBERT in all datasets, MESc achieves a significant increase in performance by at least 
    4 points in all metrics than their fine-tuned LLM counterparts with just the last layer. Combining the last four layers in 1 $\times$ encoder yields a performance boost of 4 points or more in ECtHR datasets while there is not much improvement in ILDC and SCOTUS. With the approximated structure labels, there is a slight performance increase in the test set of ILDC with $\approx$ 1 point increase in the validation set. The same goes for SCOTUS with $\approx$ 1 point increase in its validation and test set. With the same configuration and 2 $\times$ encoder, 
    we can see a much bigger performance with the structure labels achieving new baseline performance in ECtHR (A) and ECtHR (B), and ILDC datasets. For SCOTUS, this improvement from the baseline is only on the validation set. This is because of the high skew of class labels in the test dataset (for ex. label 5 has only 5 samples). With these results, we fixed 
    certain parameters in MESc for further experiments with the extracted embeddings from GPT-Neo and GPT-J. For them, we ran experiments with 2 $\times$ encoders and the last layer 
    and gained lesser performance than its 2 (or 4) layers with 2 $\times$ encoders, which we exclude in this paper. For LexGLUE, as can be seen in the table \ref{results_table_lexglue}, concatenating the embeddings from the last two layers of GPT-Neo or GPT-J had a significant impact above their vanilla fine-tuned variants by a minimum margin of 3 points for GPT-Neo-1.3B, and 1 point for GPT-Neo-2.7B and GPT-J. This increases further by a minimum of 1 point when including the approximated structure labels, showing the importance of having structural information for such sparse-annotated documents. 
    For ILDC in table \ref{tab:results_MESc_ildc}, concatenating the last four layers didn't have much improvement in the performance, while including the generated structure labels in it did increase the performance by 1 point in the validation set and slightly in the test set. 
    
\begin{figure}[h]
  \centering
  \includegraphics[height=5cm, width=\columnwidth]{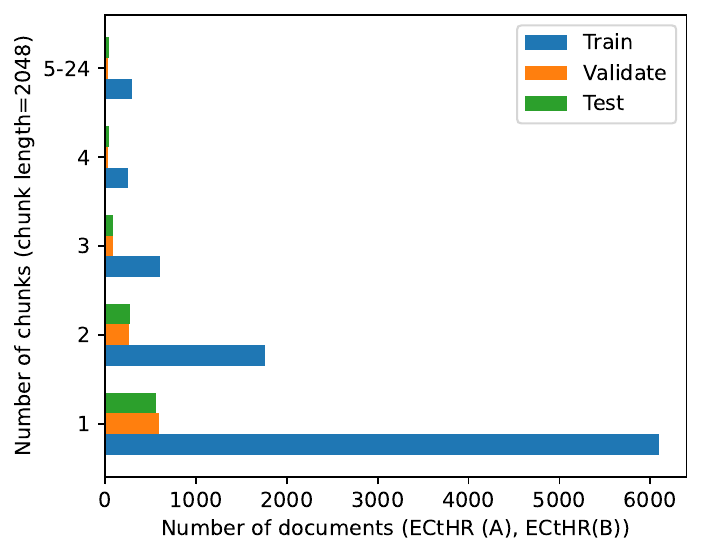} 
  \caption{Number of documents vs. the number of chunks for ECtHR.}
  \label{fig:ecthr_chunk_counts}
\end{figure}

    \paragraph{\textbf{Encoders fine-tuned on 2048 input length $(\gamma)$:}} For the sparse-annotated documents in LexGLUE and ILDC, we did a comparative study of MESc(on GPT-J 6B* ($\gamma$))'s performance with its backbone fine-tuned encoder(GPT-J 6B ($\gamma$)) (Table \ref{tab:finetuning_result}, \ref{results_table_lexglue}) to see the effect of increasing the number of parameters and the input length. GPT-J 6B ($\gamma$) fine-tuned on its maximum input length (2048) achieves better (or similar) performance than its MESc overhead trained on its extracted embeddings. For SCOTUS, MESc achieves better performance (2 points, m-F1) in the validation set but lower (2 points, m-F1) in the test set. Almost similar performance (m-F1) in ECtHR(B), 1 point higher (m-F1) in ECtHR(A)'s test set, and lesser in ILDC. 
    To check if this is the case with GPT-Neo-1.b and GPT-Neo-2.7B we fine-tuned them with their maximum input length (2048) on SCOTUS (which through our experiments can be seen as 
    more difficult to classify).
    We found for GPT-Neo (1.3B and 2.7B) fine-tuning on their maximum input length didn't show the same results as with the GPT-J, where we can see 
    that for both GPT-Neo-1.3B($\gamma$) and GPT-Neo-2.7B($\gamma$) even the MESc (on GPT-Neo-1.3B($\alpha$)) and MESc (on GPT-Neo-2.7B($\alpha$)) performs better (> 1 point m-F1) respectively. 
    To analyze this, we plot the distribution of the number of documents with respect to their chunk counts (chunk length = 2048) in the datasets 
    , one such example of ECtHR can be found in figure \ref{fig:ecthr_chunk_counts}. 
    As observed, most of the documents are able to fit between 
    1-2 chunks (median = 1), which means that with the longer input of 2048, most 
    of the important information is not fragmented during the fine-tuning process of stage 1 and is learned together. Along with this, the higher number of parameters in GPT-J is 
    able to adapt better to most of the documents. 
    We observe that most (> 90\%) of the documents can fit in very few chunks, deepening the models with extra layers (stages 3 \& 4) does not have any added value. 
    

    With these results, we find that: 
    \begin{enumerate}
        \item  
        Concatenating the last two layers in GPT-Neo (1.3b,2.7B) or GPT-J provides the optimum number of feature variances. And for BERT-based models, the last 4 layers worked better. 
        Globally concatenating the embeddings helped to get a better approximation of the structure labels and improves the performance.
        \item MESc adapts well to LLMs (BERT-based models, GPT-Neo-1.3B, GPT-Neo-2.7B) with less than 6 billion parameters (GPT-J).
        \item MESc works better than its counterpart LLM under the condition that the length of most of the documents in the dataset is much greater than the maximum input length of the LLM. 
    \end{enumerate}
\subsection{Results on extractive explanation (ORSE)}

\begin{table}[h] \footnotesize
\caption{ORSE vs ILDC$_{expert}$ \cite{ildc-cjpe}}
\label{tab:Explanation_res}
\begin{tabular}{|l|c|c|c|c|c|} \cline{2-6}
\multicolumn{1}{l|}{} & \multicolumn{5}{c|}{Expert} \\ \cline{2-6}
\multicolumn{1}{l|}{} & 1 & 2 & 3 & 4 & 5 \\ \cline{2-6}
\multicolumn{1}{l|}{} & \multicolumn{5}{c|}{Baseline-scores} \\ \hline
\textit{ROUGE-1} & 0.444 & 0.517 & 0.401 & 0.391 & 0.501 \\ \hline
\textit{ROUGE-2} & 0.303 & 0.295 & 0.296 & 0.297 & 0.294 \\ \hline
\textit{ROUGE-L} & 0.439 & 0.407 & 0.423 & 0.444 & 0.407 \\ \hline
\textit{BLEU} & 0.16 & 0.28 & 0.099 & 0.093 & 0.248 \\ \hline
\textit{Jaccard Similarity} & 0.333 & 0.317 & 0.328 & 0.324 & 0.318 \\ \hline
\multicolumn{1}{l|}{} & \multicolumn{5}{c|}{ORSE @ k=20\% (MESc with InLegalBERT ($\alpha$))} \\ \hline
\textit{ROUGE-1} & 0.4844 & 0.4815 & 0.4657 & 0.4678 & 0.5083 \\ \hline
\textit{ROUGE-2} & 0.3339 & 0.3125 & 0.3299 & 0.331 & 0.3523 \\ \hline
\textit{ROUGE-L} & 0.4679 & 0.4518 & 0.4537 & 0.4577 & 0.488 \\ \hline
\textit{BLEU} & 0.1682 & 0.3253 & 0.0969 & 0.08 & 0.2973 \\ \hline
\textit{Jaccard Similarity} & 0.346 & 0.3381 & 0.3306 & 0.318 & 0.3637 \\ \hline
\multicolumn{1}{l|}{} & \multicolumn{5}{c|}{ORSE @ k=30\% (MESc with InLegalBERT ($\alpha$))} \\ \hline
\textit{ROUGE-1} & 0.5441 & 0.5108 & 0.5351 & 0.5445 & 0.55 \\ \hline
\textit{ROUGE-2} & 0.3939 & 0.3452 & 0.4018 & 0.4078 & 0.3976 \\ \hline
\textit{ROUGE-L} & 0.5266 & 0.4815 & 0.5225 & 0.5338 & 0.5302 \\ \hline
\textit{BLEU} & 0.2888 & 0.3979 & 0.2104 & 0.1901 & 0.4049 \\ \hline
\textit{Jaccard Similarity} & 0.4051 & 0.3657 & 0.3992 & 0.3896 & 0.4044 \\ \hline
\multicolumn{1}{l|}{} & \multicolumn{5}{c|}{ORSE @ k=40\% (MESc with InLegalBERT ($\alpha$))} \\ \hline
\textit{ROUGE-1} & 0.5809 & 0.5201 & 0.5857 & 0.6016 & 0.5665 \\ \hline
\textit{ROUGE-2} & 0.4364 & 0.3574 & 0.4597 & 0.4738 & 0.4185 \\ \hline
\textit{ROUGE-L} & 0.5649 & 0.4942 & 0.5741 & 0.5914 & 0.5476 \\ \hline
\textit{BLEU} & 0.3918 & 0.3915 & 0.3416 & 0.3221 & 0.4397 \\ \hline
\textit{Jaccard Similarity} & 0.4445 & 0.3739 & 0.4535 & 0.4476 & 0.4216 \\ \hline
\multicolumn{1}{l|}{} & \multicolumn{5}{c|}{ORSE @ k=30\% (MESc with GPT-J 6B ($\alpha$))} \\ \hline
\textit{ROUGE-1} & \textbf{0.5448} & \textbf{0.5152} & \textbf{0.5327} & \textbf{0.5484} & \textbf{0.5488} \\ \hline
\textit{ROUGE-2} & \textbf{0.3996} & \textbf{0.3497} & \textbf{0.4009} & \textbf{0.4159} & \textbf{0.3931} \\ \hline
\textit{ROUGE-L} & \textbf{0.5277} & \textbf{0.4858} & \textbf{0.5182} & \textbf{0.5387} & \textbf{0.5283} \\ \hline
\textit{BLEU} & \textbf{0.2834} & \textbf{0.4078} & \textbf{0.2057} & \textbf{0.1861} & \textbf{0.3982} \\ \hline
\textit{Jaccard Similarity} & \textbf{0.4076} & \textbf{0.372} & \textbf{0.3984} & \textbf{0.3948} & \textbf{0.4057} \\ \hline
\multicolumn{1}{l|}{} & \multicolumn{5}{c|}{ORSE @ k=40\% (MESc with GPT-J 6B ($\alpha$))} \\ \hline
\textit{ROUGE-1} & \textbf{0.5822} & \textbf{0.5297} & \textbf{0.5864} & \textbf{0.6077} & \textbf{0.567} \\ \hline
\textit{ROUGE-2} & \textbf{0.4414} & \textbf{0.3685} & \textbf{0.4611} & \textbf{0.4816} & \textbf{0.4164} \\ \hline
\textit{ROUGE-L} & \textbf{0.5659} & \textbf{0.5033} & \textbf{0.573} & \textbf{0.5984} & \textbf{0.5464} \\ \hline
\textit{BLEU} & \textbf{0.3854} & \textbf{0.4082} & \textbf{0.3288} & \textbf{0.3113} & \textbf{0.4305} \\ \hline
\textit{Jaccard Similarity} & \textbf{0.4479} & \textbf{0.3854} & \textbf{0.4563} & \textbf{0.4566} & \textbf{0.4246} \\ \hline
\end{tabular}
\end{table}
We used two best-performing configurations of MESc in ILDC (InLegalBERT ($\alpha$) and GPT-J 6B ($\alpha$)) to extract the sentences with ORSE and varying k as 20\%, 30\%, and 40\%.
The performance of ORSE can be seen in table \ref{tab:Explanation_res}. 
The sentences extracted from our extractive explanation algorithm are compared with the gold explanations given by five different annotators (1,2,3,4,5) in ILDC$_{Expert}$. The sentence similarity between the two is measured with the help of metrics ROUGE-1, ROUGE-2, ROUGE-L \cite{rouge}, Jaccard similarity, and BLEU\cite{bleu} score, the results of which are shown in Table \ref{tab:Explanation_res}. We compare scores from our algorithm with the baseline score on the ILDC$_{Expert }$ \cite{ildc-cjpe}.
With InLegalBERT ($\alpha$), and k = top 20\% of ranked sentences, ORSE performs almost similar to the baseline in ROUGE-1 while overall slightly better in other metrics. With k = 30\%, ORSE surpasses the baselines with a total gain of 19\% on ROUGE-1, 31\% on ROUGE-2, 22.38\% on ROUGE-L, 69.5\% on BLEU, and 32.16\% on Jaccard Similarity. For k = 40\% the gain is much higher with a total gain of 26.65\% on ROUGE-1, 44.49\% on ROUGE-2, 30.76\% on ROUGE-L, 114\% on BLEU, and 32.16\% on Jaccard Similarity. The explanations extracted from GPT-J 6B ($\alpha$) variant are slightly better than InLegalBERT ($\alpha$) for both k = 30\% and 40\% respectively. Overall ORSE performs better than the baseline with the best scores having a total average gain of 50\% over the baseline on all the metrics.

\section{Conclusion}
We explore the problem of classification of large and unstructured legal documents and develop a multi-stage hierarchical classification framework (MESc). 
We find the effect of including the structure information with our approximated structure labels in such documents and also explore the impact of combining the embeddings from the last layers of a fine-tuned transformer encoder model in MESc.
Along with BERT-based LLMs, we also explored the adaptability of larger LLMs (GPT-Neo and GPT-J) with multi-billion parameters, to MESc. 
We check MESc's 
limits (section \ref{Performance with MESc}) with these LLMs to suggest 
the optimal condition for its performance. 
GPT-Neo and GPT-J adapted well to legal cases from India and Europe even though they were pre-trained only on the US legal case documents showing the intra-domain(legal) transfer learning capacity of these multi-billion parameter language models. 
Our experiments achieve a new benchmark in the classification of the ILDC and the LexGLUE subset (ECtHR (A), ECtHR (B), and SCOTUS). 
For the explanation in such hierarchical models, we developed an extractive explanation algorithm (ORSE) based on the sensitivity of a model to its inputs at each level of the hierarchy. ORSE ranks the sentences according to their impact on the prediction/classification and achieves an average performance gain of 50\% in ILDC$_{Expert}$ over the previous benchmark. 
We aim to further develop the explanation algorithm to adapt to a general neural framework in our future work. Alongside we also aim to leverage this work in-domain, on the French and European legal cases by exploring further the problem of length and non-uniform structure in these legal case documents. 

\section{Ethical Considerations}
\label{ethical}
Our work aligns with the ethical consideration of the datasets (ILDC \cite{ildc-cjpe} and LexGLUE \cite{lexglue})) used here for 
the experimentation and evaluation of our approach. We add certain points to this. The framework developed here is in no way to create a "robotic" judge or replace one in real life. Rather we try to create such frameworks to analyze how deep learning and natural language processing techniques can be applied to legal documents to extract and provide legal professionals with patterns and insights that may not be implicitly 
visible. 
The 
methods developed here are in no way foolproof to predict and generate an explanatory response, and should not be used for the same in real-life settings (courts) or used to guide people unfamiliar with legal proceedings. The results from our framework should not be used by a non-professional to make high-stakes decisions in one's life concerning legal cases.

\begin{acks}
This work is supported by the LAWBOT project (ANR-20-CE38-0013) and HPC/AI resources from GENCI-IDRIS grant number 2022-AD011013937.
\end{acks}
\bibliographystyle{ACM-Reference-Format}
\bibliography{MESc}


\begin{thebibliography}{47}


\ifx \showCODEN    \undefined \def \showCODEN     #1{\unskip}     \fi
\ifx \showDOI      \undefined \def \showDOI       #1{#1}\fi
\ifx \showISBNx    \undefined \def \showISBNx     #1{\unskip}     \fi
\ifx \showISBNxiii \undefined \def \showISBNxiii  #1{\unskip}     \fi
\ifx \showISSN     \undefined \def \showISSN      #1{\unskip}     \fi
\ifx \showLCCN     \undefined \def \showLCCN      #1{\unskip}     \fi
\ifx \shownote     \undefined \def \shownote      #1{#1}          \fi
\ifx \showarticletitle \undefined \def \showarticletitle #1{#1}   \fi
\ifx \showURL      \undefined \def \showURL       {\relax}        \fi
\providecommand\bibfield[2]{#2}
\providecommand\bibinfo[2]{#2}
\providecommand\natexlab[1]{#1}
\providecommand\showeprint[2][]{arXiv:#2}

\bibitem[Ainslie et~al\mbox{.}(2020)]%
        {ETC_transformer}
\bibfield{author}{\bibinfo{person}{Joshua Ainslie}, \bibinfo{person}{Santiago
  Ontanon}, \bibinfo{person}{Chris Alberti}, \bibinfo{person}{Vaclav Cvicek},
  \bibinfo{person}{Zachary Fisher}, \bibinfo{person}{Philip Pham},
  \bibinfo{person}{Anirudh Ravula}, \bibinfo{person}{Sumit Sanghai},
  \bibinfo{person}{Qifan Wang}, {and} \bibinfo{person}{Li Yang}.}
  \bibinfo{year}{2020}\natexlab{}.
\newblock \showarticletitle{{ETC}: Encoding Long and Structured Inputs in
  Transformers}. In \bibinfo{booktitle}{\emph{Proceedings of the 2020
  Conference on Empirical Methods in Natural Language Processing (EMNLP)}}.
  \bibinfo{publisher}{Association for Computational Linguistics},
  \bibinfo{address}{Online}, \bibinfo{pages}{268--284}.
\newblock
\urldef\tempurl%
\url{https://doi.org/10.18653/v1/2020.emnlp-main.19}
\showDOI{\tempurl}


\bibitem[Beltagy et~al\mbox{.}(2020)]%
        {longformer}
\bibfield{author}{\bibinfo{person}{Iz Beltagy}, \bibinfo{person}{Matthew~E.
  Peters}, {and} \bibinfo{person}{Arman Cohan}.}
  \bibinfo{year}{2020}\natexlab{}.
\newblock \bibinfo{title}{Longformer: The Long-Document Transformer}.
\newblock
\newblock
\urldef\tempurl%
\url{https://doi.org/10.48550/ARXIV.2004.05150}
\showDOI{\tempurl}


\bibitem[Black et~al\mbox{.}(2021)]%
        {gpt-neo}
\bibfield{author}{\bibinfo{person}{Sid Black}, \bibinfo{person}{Gao Leo},
  \bibinfo{person}{Phil Wang}, \bibinfo{person}{Connor Leahy}, {and}
  \bibinfo{person}{Stella Biderman}.} \bibinfo{year}{2021}\natexlab{}.
\newblock \bibinfo{booktitle}{\emph{{GPT-Neo: Large Scale Autoregressive
  Language Modeling with Mesh-Tensorflow}}}.
\newblock
\urldef\tempurl%
\url{https://doi.org/10.5281/zenodo.5297715}
\showDOI{\tempurl}


\bibitem[Brown et~al\mbox{.}(2020)]%
        {GPT-3}
\bibfield{author}{\bibinfo{person}{Tom Brown}, \bibinfo{person}{Benjamin Mann},
  \bibinfo{person}{Nick Ryder}, \bibinfo{person}{Melanie Subbiah},
  \bibinfo{person}{Jared~D Kaplan}, \bibinfo{person}{Prafulla Dhariwal},
  \bibinfo{person}{Arvind Neelakantan}, \bibinfo{person}{Pranav Shyam},
  \bibinfo{person}{Girish Sastry}, \bibinfo{person}{Amanda Askell},
  \bibinfo{person}{Sandhini Agarwal}, \bibinfo{person}{Ariel Herbert-Voss},
  \bibinfo{person}{Gretchen Krueger}, \bibinfo{person}{Tom Henighan},
  \bibinfo{person}{Rewon Child}, \bibinfo{person}{Aditya Ramesh},
  \bibinfo{person}{Daniel Ziegler}, \bibinfo{person}{Jeffrey Wu},
  \bibinfo{person}{Clemens Winter}, \bibinfo{person}{Chris Hesse},
  \bibinfo{person}{Mark Chen}, \bibinfo{person}{Eric Sigler},
  \bibinfo{person}{Mateusz Litwin}, \bibinfo{person}{Scott Gray},
  \bibinfo{person}{Benjamin Chess}, \bibinfo{person}{Jack Clark},
  \bibinfo{person}{Christopher Berner}, \bibinfo{person}{Sam McCandlish},
  \bibinfo{person}{Alec Radford}, \bibinfo{person}{Ilya Sutskever}, {and}
  \bibinfo{person}{Dario Amodei}.} \bibinfo{year}{2020}\natexlab{}.
\newblock \showarticletitle{Language Models are Few-Shot Learners}. In
  \bibinfo{booktitle}{\emph{Advances in Neural Information Processing
  Systems}}, \bibfield{editor}{\bibinfo{person}{H.~Larochelle},
  \bibinfo{person}{M.~Ranzato}, \bibinfo{person}{R.~Hadsell},
  \bibinfo{person}{M.F. Balcan}, {and} \bibinfo{person}{H.~Lin}} (Eds.),
  Vol.~\bibinfo{volume}{33}. \bibinfo{publisher}{Curran Associates, Inc.},
  \bibinfo{pages}{1877--1901}.
\newblock
\urldef\tempurl%
\url{https://proceedings.neurips.cc/paper_files/paper/2020/file/1457c0d6bfcb4967418bfb8ac142f64a-Paper.pdf}
\showURL{%
\tempurl}


\bibitem[Chalkidis et~al\mbox{.}(2019)]%
        {etchr_a}
\bibfield{author}{\bibinfo{person}{Ilias Chalkidis}, \bibinfo{person}{Ion
  Androutsopoulos}, {and} \bibinfo{person}{Nikolaos Aletras}.}
  \bibinfo{year}{2019}\natexlab{}.
\newblock \showarticletitle{Neural Legal Judgment Prediction in {E}nglish}. In
  \bibinfo{booktitle}{\emph{Proceedings of the 57th Annual Meeting of the
  Association for Computational Linguistics}}. \bibinfo{publisher}{Association
  for Computational Linguistics}, \bibinfo{address}{Florence, Italy},
  \bibinfo{pages}{4317--4323}.
\newblock
\urldef\tempurl%
\url{https://doi.org/10.18653/v1/P19-1424}
\showDOI{\tempurl}


\bibitem[Chalkidis et~al\mbox{.}(2022a)]%
        {HAT}
\bibfield{author}{\bibinfo{person}{Ilias Chalkidis}, \bibinfo{person}{Xiang
  Dai}, \bibinfo{person}{Manos Fergadiotis}, \bibinfo{person}{Prodromos
  Malakasiotis}, {and} \bibinfo{person}{Desmond Elliott}.}
  \bibinfo{year}{2022}\natexlab{a}.
\newblock \bibinfo{title}{An Exploration of Hierarchical Attention Transformers
  for Efficient Long Document Classification}.
\newblock
\newblock
\urldef\tempurl%
\url{https://arxiv.org/abs/2210.05529}
\showURL{%
\tempurl}


\bibitem[Chalkidis et~al\mbox{.}(2020)]%
        {legal-bert}
\bibfield{author}{\bibinfo{person}{Ilias Chalkidis}, \bibinfo{person}{Manos
  Fergadiotis}, \bibinfo{person}{Prodromos Malakasiotis},
  \bibinfo{person}{Nikolaos Aletras}, {and} \bibinfo{person}{Ion
  Androutsopoulos}.} \bibinfo{year}{2020}\natexlab{}.
\newblock \showarticletitle{{LEGAL}-{BERT}: The Muppets straight out of Law
  School}. In \bibinfo{booktitle}{\emph{Findings of the Association for
  Computational Linguistics: EMNLP 2020}}. \bibinfo{publisher}{Association for
  Computational Linguistics}, \bibinfo{address}{Online},
  \bibinfo{pages}{2898--2904}.
\newblock
\urldef\tempurl%
\url{https://doi.org/10.18653/v1/2020.findings-emnlp.261}
\showDOI{\tempurl}


\bibitem[Chalkidis et~al\mbox{.}(2021)]%
        {ecthr_b}
\bibfield{author}{\bibinfo{person}{Ilias Chalkidis}, \bibinfo{person}{Manos
  Fergadiotis}, \bibinfo{person}{Dimitrios Tsarapatsanis},
  \bibinfo{person}{Nikolaos Aletras}, \bibinfo{person}{Ion Androutsopoulos},
  {and} \bibinfo{person}{Prodromos Malakasiotis}.}
  \bibinfo{year}{2021}\natexlab{}.
\newblock \showarticletitle{Paragraph-level Rationale Extraction through
  Regularization: A case study on {E}uropean Court of Human Rights Cases}. In
  \bibinfo{booktitle}{\emph{Proceedings of the 2021 Conference of the North
  American Chapter of the Association for Computational Linguistics: Human
  Language Technologies}}. \bibinfo{publisher}{Association for Computational
  Linguistics}, \bibinfo{address}{Online}, \bibinfo{pages}{226--241}.
\newblock
\urldef\tempurl%
\url{https://doi.org/10.18653/v1/2021.naacl-main.22}
\showDOI{\tempurl}


\bibitem[Chalkidis et~al\mbox{.}(2022b)]%
        {lexglue}
\bibfield{author}{\bibinfo{person}{Ilias Chalkidis}, \bibinfo{person}{Abhik
  Jana}, \bibinfo{person}{Dirk Hartung}, \bibinfo{person}{Michael Bommarito},
  \bibinfo{person}{Ion Androutsopoulos}, \bibinfo{person}{Daniel Katz}, {and}
  \bibinfo{person}{Nikolaos Aletras}.} \bibinfo{year}{2022}\natexlab{b}.
\newblock \showarticletitle{{L}ex{GLUE}: A Benchmark Dataset for Legal Language
  Understanding in {E}nglish}. In \bibinfo{booktitle}{\emph{Proceedings of the
  60th Annual Meeting of the Association for Computational Linguistics (Volume
  1: Long Papers)}}. \bibinfo{publisher}{Association for Computational
  Linguistics}, \bibinfo{address}{Dublin, Ireland},
  \bibinfo{pages}{4310--4330}.
\newblock
\urldef\tempurl%
\url{https://doi.org/10.18653/v1/2022.acl-long.297}
\showDOI{\tempurl}


\bibitem[Chen et~al\mbox{.}(2019)]%
        {Huajie}
\bibfield{author}{\bibinfo{person}{Huajie Chen}, \bibinfo{person}{Deng Cai},
  \bibinfo{person}{Wei Dai}, \bibinfo{person}{Zehui Dai}, {and}
  \bibinfo{person}{Yadong Ding}.} \bibinfo{year}{2019}\natexlab{}.
\newblock \showarticletitle{Charge-Based Prison Term Prediction with Deep
  Gating Network}. In \bibinfo{booktitle}{\emph{Proceedings of the 2019
  Conference on Empirical Methods in Natural Language Processing and the 9th
  International Joint Conference on Natural Language Processing
  (EMNLP-IJCNLP)}}. \bibinfo{publisher}{Association for Computational
  Linguistics}, \bibinfo{address}{Hong Kong, China},
  \bibinfo{pages}{6362--6367}.
\newblock
\urldef\tempurl%
\url{https://doi.org/10.18653/v1/D19-1667}
\showDOI{\tempurl}


\bibitem[Cui et~al\mbox{.}(2022)]%
        {survey_2}
\bibfield{author}{\bibinfo{person}{Junyun Cui}, \bibinfo{person}{Xiaoyu Shen},
  \bibinfo{person}{Feiping Nie}, \bibinfo{person}{Z. Wang},
  \bibinfo{person}{Jinglong Wang}, {and} \bibinfo{person}{Yulong Chen}.}
  \bibinfo{year}{2022}\natexlab{}.
\newblock \showarticletitle{A Survey on Legal Judgment Prediction: Datasets,
  Metrics, Models and Challenges}.
\newblock \bibinfo{journal}{\emph{ArXiv}}  \bibinfo{volume}{abs/2204.04859}
  (\bibinfo{year}{2022}).
\newblock


\bibitem[Devlin et~al\mbox{.}(2019)]%
        {bert}
\bibfield{author}{\bibinfo{person}{Jacob Devlin}, \bibinfo{person}{Ming{-}Wei
  Chang}, \bibinfo{person}{Kenton Lee}, {and} \bibinfo{person}{Kristina
  Toutanova}.} \bibinfo{year}{2019}\natexlab{}.
\newblock \showarticletitle{{BERT:} Pre-training of Deep Bidirectional
  Transformers for Language Understanding}. In
  \bibinfo{booktitle}{\emph{Proceedings of the 2019 Conference of the North
  American Chapter of the Association for Computational Linguistics: Human
  Language Technologies, {NAACL-HLT} 2019, Minneapolis, MN, USA, June 2-7,
  2019, Volume 1 (Long and Short Papers)}},
  \bibfield{editor}{\bibinfo{person}{Jill Burstein}, \bibinfo{person}{Christy
  Doran}, {and} \bibinfo{person}{Thamar Solorio}} (Eds.).
  \bibinfo{publisher}{Association for Computational Linguistics},
  \bibinfo{pages}{4171--4186}.
\newblock
\urldef\tempurl%
\url{https://doi.org/10.18653/v1/n19-1423}
\showDOI{\tempurl}


\bibitem[Feng et~al\mbox{.}(2022)]%
        {survey_1}
\bibfield{author}{\bibinfo{person}{Yi Feng}, \bibinfo{person}{Chuanyi Li},
  {and} \bibinfo{person}{Vincent Ng}.} \bibinfo{year}{2022}\natexlab{}.
\newblock \showarticletitle{Legal Judgment Prediction: A Survey of the State of
  the Art}. In \bibinfo{booktitle}{\emph{Proceedings of the Thirty-First
  International Joint Conference on Artificial Intelligence, {IJCAI-22}}},
  \bibfield{editor}{\bibinfo{person}{Lud~De Raedt}} (Ed.).
  \bibinfo{publisher}{International Joint Conferences on Artificial
  Intelligence Organization}, \bibinfo{pages}{5461--5469}.
\newblock
\urldef\tempurl%
\url{https://doi.org/10.24963/ijcai.2022/765}
\showDOI{\tempurl}
\newblock
\shownote{Survey Track}.


\bibitem[Gao et~al\mbox{.}(2021)]%
        {pile}
\bibfield{author}{\bibinfo{person}{Leo Gao}, \bibinfo{person}{Stella Biderman},
  \bibinfo{person}{Sid Black}, \bibinfo{person}{Laurence Golding},
  \bibinfo{person}{Travis Hoppe}, \bibinfo{person}{Charles Foster},
  \bibinfo{person}{Jason Phang}, \bibinfo{person}{Horace He},
  \bibinfo{person}{Anish Thite}, \bibinfo{person}{Noa Nabeshima},
  \bibinfo{person}{Shawn Presser}, {and} \bibinfo{person}{Connor Leahy}.}
  \bibinfo{year}{2021}\natexlab{}.
\newblock \showarticletitle{The Pile: An 800GB Dataset of Diverse Text for
  Language Modeling}.
\newblock \bibinfo{journal}{\emph{CoRR}}  \bibinfo{volume}{abs/2101.00027}
  (\bibinfo{year}{2021}).
\newblock
\showeprint[arXiv]{2101.00027}
\urldef\tempurl%
\url{https://arxiv.org/abs/2101.00027}
\showURL{%
\tempurl}


\bibitem[Jiang et~al\mbox{.}(2018)]%
        {jiang-etal-interpretable}
\bibfield{author}{\bibinfo{person}{Xin Jiang}, \bibinfo{person}{Hai Ye},
  \bibinfo{person}{Zhunchen Luo}, \bibinfo{person}{WenHan Chao}, {and}
  \bibinfo{person}{Wenjia Ma}.} \bibinfo{year}{2018}\natexlab{}.
\newblock \showarticletitle{Interpretable Rationale Augmented Charge Prediction
  System}. In \bibinfo{booktitle}{\emph{Proceedings of the 27th International
  Conference on Computational Linguistics: System Demonstrations}}.
  \bibinfo{publisher}{Association for Computational Linguistics},
  \bibinfo{address}{Santa Fe, New Mexico}, \bibinfo{pages}{146--151}.
\newblock
\urldef\tempurl%
\url{https://aclanthology.org/C18-2032}
\showURL{%
\tempurl}


\bibitem[Katju(2019)]%
        {katju_pending_cases}
\bibfield{author}{\bibinfo{person}{Justice~Markandey Katju}.}
  \bibinfo{year}{2019}\natexlab{}.
\newblock \showarticletitle{Backlog of cases crippling judiciary}.
\newblock  (\bibinfo{year}{2019}).
\newblock
\urldef\tempurl%
\url{https://www.tribuneindia.com/news/archive/comment/backlog-of-cases-crippling-judiciary-776503}
\showURL{%
\tempurl}


\bibitem[Kaufman et~al\mbox{.}(2019)]%
        {usa_kaufman_kraft}
\bibfield{author}{\bibinfo{person}{Aaron~Russell Kaufman},
  \bibinfo{person}{Peter Kraft}, {and} \bibinfo{person}{Maya Sen}.}
  \bibinfo{year}{2019}\natexlab{}.
\newblock \showarticletitle{Improving Supreme Court Forecasting Using Boosted
  Decision Trees}.
\newblock \bibinfo{journal}{\emph{Political Analysis}} \bibinfo{volume}{27},
  \bibinfo{number}{3} (\bibinfo{year}{2019}), \bibinfo{pages}{381–387}.
\newblock
\urldef\tempurl%
\url{https://doi.org/10.1017/pan.2018.59}
\showDOI{\tempurl}


\bibitem[Kitaev et~al\mbox{.}(2020)]%
        {reformer}
\bibfield{author}{\bibinfo{person}{Nikita Kitaev}, \bibinfo{person}{Lukasz
  Kaiser}, {and} \bibinfo{person}{Anselm Levskaya}.}
  \bibinfo{year}{2020}\natexlab{}.
\newblock \showarticletitle{Reformer: The Efficient Transformer}. In
  \bibinfo{booktitle}{\emph{8th International Conference on Learning
  Representations, {ICLR} 2020, Addis Ababa, Ethiopia, April 26-30, 2020}}.
  \bibinfo{publisher}{OpenReview.net}.
\newblock
\urldef\tempurl%
\url{https://openreview.net/forum?id=rkgNKkHtvB}
\showURL{%
\tempurl}


\bibitem[Lin(2004)]%
        {rouge}
\bibfield{author}{\bibinfo{person}{Chin-Yew Lin}.}
  \bibinfo{year}{2004}\natexlab{}.
\newblock \showarticletitle{{ROUGE}: A Package for Automatic Evaluation of
  Summaries}. In \bibinfo{booktitle}{\emph{Text Summarization Branches Out}}.
  \bibinfo{publisher}{Association for Computational Linguistics},
  \bibinfo{address}{Barcelona, Spain}, \bibinfo{pages}{74--81}.
\newblock
\urldef\tempurl%
\url{https://aclanthology.org/W04-1013}
\showURL{%
\tempurl}


\bibitem[Malik et~al\mbox{.}(2021)]%
        {ildc-cjpe}
\bibfield{author}{\bibinfo{person}{Vijit Malik}, \bibinfo{person}{Rishabh
  Sanjay}, \bibinfo{person}{Shubham~Kumar Nigam}, \bibinfo{person}{Kripa
  Ghosh}, \bibinfo{person}{Shouvik~Kumar Guha}, \bibinfo{person}{Arnab
  Bhattacharya}, {and} \bibinfo{person}{Ashutosh Modi}.}
  \bibinfo{year}{2021}\natexlab{}.
\newblock \showarticletitle{{ILDC} for {CJPE:} Indian Legal Documents Corpus
  for Court JudgmentPrediction and Explanation}.
\newblock \bibinfo{journal}{\emph{CoRR}}  \bibinfo{volume}{abs/2105.13562}
  (\bibinfo{year}{2021}).
\newblock
\showeprint[arXiv]{2105.13562}
\urldef\tempurl%
\url{https://arxiv.org/abs/2105.13562}
\showURL{%
\tempurl}


\bibitem[McInnes et~al\mbox{.}(2017)]%
        {hdbscan}
\bibfield{author}{\bibinfo{person}{Leland McInnes}, \bibinfo{person}{John
  Healy}, {and} \bibinfo{person}{Steve Astels}.}
  \bibinfo{year}{2017}\natexlab{}.
\newblock \showarticletitle{hdbscan: Hierarchical density based clustering}.
\newblock \bibinfo{journal}{\emph{Journal of Open Source Software}}
  \bibinfo{volume}{2}, \bibinfo{number}{11} (\bibinfo{year}{2017}),
  \bibinfo{pages}{205}.
\newblock
\urldef\tempurl%
\url{https://doi.org/10.21105/joss.00205}
\showDOI{\tempurl}


\bibitem[McInnes et~al\mbox{.}(2018)]%
        {umap}
\bibfield{author}{\bibinfo{person}{Leland McInnes}, \bibinfo{person}{John
  Healy}, {and} \bibinfo{person}{James Melville}.}
  \bibinfo{year}{2018}\natexlab{}.
\newblock \bibinfo{title}{UMAP: Uniform Manifold Approximation and Projection
  for Dimension Reduction}.
\newblock
\newblock
\urldef\tempurl%
\url{https://doi.org/10.48550/ARXIV.1802.03426}
\showDOI{\tempurl}


\bibitem[Nallapati and Manning(2008)]%
        {diff_from_generalTxt_2}
\bibfield{author}{\bibinfo{person}{Ramesh Nallapati} {and}
  \bibinfo{person}{Christopher~D. Manning}.} \bibinfo{year}{2008}\natexlab{}.
\newblock \showarticletitle{Legal Docket Classification: {W}here Machine
  Learning Stumbles}. In \bibinfo{booktitle}{\emph{Proceedings of the 2008
  Conference on Empirical Methods in Natural Language Processing}}.
  \bibinfo{publisher}{Association for Computational Linguistics},
  \bibinfo{address}{Honolulu, Hawaii}, \bibinfo{pages}{438--446}.
\newblock
\urldef\tempurl%
\url{https://aclanthology.org/D08-1046}
\showURL{%
\tempurl}


\bibitem[Papineni et~al\mbox{.}(2002)]%
        {bleu}
\bibfield{author}{\bibinfo{person}{Kishore Papineni}, \bibinfo{person}{Salim
  Roukos}, \bibinfo{person}{Todd Ward}, {and} \bibinfo{person}{Wei-Jing Zhu}.}
  \bibinfo{year}{2002}\natexlab{}.
\newblock \showarticletitle{{B}leu: a Method for Automatic Evaluation of
  Machine Translation}. In \bibinfo{booktitle}{\emph{Proceedings of the 40th
  Annual Meeting of the Association for Computational Linguistics}}.
  \bibinfo{publisher}{Association for Computational Linguistics},
  \bibinfo{address}{Philadelphia, Pennsylvania, USA},
  \bibinfo{pages}{311--318}.
\newblock
\urldef\tempurl%
\url{https://doi.org/10.3115/1073083.1073135}
\showDOI{\tempurl}


\bibitem[Pappagari et~al\mbox{.}(2019)]%
        {Hierarchical_Transformers}
\bibfield{author}{\bibinfo{person}{Raghavendra Pappagari},
  \bibinfo{person}{Piotr Zelasko}, \bibinfo{person}{Jesús Villalba},
  \bibinfo{person}{Yishay Carmiel}, {and} \bibinfo{person}{Najim Dehak}.}
  \bibinfo{year}{2019}\natexlab{}.
\newblock \showarticletitle{Hierarchical Transformers for Long Document
  Classification}. In \bibinfo{booktitle}{\emph{2019 IEEE Automatic Speech
  Recognition and Understanding Workshop (ASRU)}}. \bibinfo{pages}{838--844}.
\newblock
\urldef\tempurl%
\url{https://doi.org/10.1109/ASRU46091.2019.9003958}
\showDOI{\tempurl}


\bibitem[Paul et~al\mbox{.}(2022)]%
        {InLegalBERT}
\bibfield{author}{\bibinfo{person}{Shounak Paul}, \bibinfo{person}{Arpan
  Mandal}, \bibinfo{person}{Pawan Goyal}, {and} \bibinfo{person}{Saptarshi
  Ghosh}.} \bibinfo{year}{2022}\natexlab{}.
\newblock \bibinfo{title}{Pre-training Transformers on Indian Legal Text}.
\newblock
\newblock
\urldef\tempurl%
\url{https://doi.org/10.48550/ARXIV.2209.06049}
\showDOI{\tempurl}


\bibitem[Petsiuk et~al\mbox{.}(2018)]%
        {rise}
\bibfield{author}{\bibinfo{person}{Vitali Petsiuk}, \bibinfo{person}{Abir Das},
  {and} \bibinfo{person}{Kate Saenko}.} \bibinfo{year}{2018}\natexlab{}.
\newblock \showarticletitle{{RISE:} Randomized Input Sampling for Explanation
  of Black-box Models}. In \bibinfo{booktitle}{\emph{British Machine Vision
  Conference 2018, {BMVC} 2018, Newcastle, UK, September 3-6, 2018}}.
  \bibinfo{publisher}{{BMVA} Press}, \bibinfo{pages}{151}.
\newblock
\urldef\tempurl%
\url{http://bmvc2018.org/contents/papers/1064.pdf}
\showURL{%
\tempurl}


\bibitem[Prasad et~al\mbox{.}(2022)]%
        {CIRCLE22}
\bibfield{author}{\bibinfo{person}{Nishchal Prasad}, \bibinfo{person}{Mohand
  Boughanem}, {and} \bibinfo{person}{Taoufiq Dkaki}.}
  \bibinfo{year}{2022}\natexlab{}.
\newblock \showarticletitle{Effect of Hierarchical Domain-specific Language
  Models and Attention in the Classification of Decisions for Legal Cases}. In
  \bibinfo{booktitle}{\emph{Proceedings of the 2nd Joint Conference of the
  Information Retrieval Communities in Europe {(CIRCLE} 2022), Samatan, Gers,
  France, July 4-7, 2022}} \emph{(\bibinfo{series}{{CEUR} Workshop
  Proceedings}, Vol.~\bibinfo{volume}{3178})}.
  \bibinfo{publisher}{CEUR-WS.org}.
\newblock
\urldef\tempurl%
\url{http://ceur-ws.org/Vol-3178/CIRCLE\_2022\_paper\_21.pdf}
\showURL{%
\tempurl}


\bibitem[Tay et~al\mbox{.}(2022)]%
        {Tay_Scaling_Laws}
\bibfield{author}{\bibinfo{person}{Yi Tay}, \bibinfo{person}{Mostafa Dehghani},
  \bibinfo{person}{Samira Abnar}, \bibinfo{person}{Hyung~Won Chung},
  \bibinfo{person}{William Fedus}, \bibinfo{person}{Jinfeng Rao},
  \bibinfo{person}{Sharan Narang}, \bibinfo{person}{Vinh~Q. Tran},
  \bibinfo{person}{Dani Yogatama}, {and} \bibinfo{person}{Donald Metzler}.}
  \bibinfo{year}{2022}\natexlab{}.
\newblock \bibinfo{title}{Scaling Laws vs Model Architectures: How does
  Inductive Bias Influence Scaling?}
\newblock
\newblock
\urldef\tempurl%
\url{https://doi.org/10.48550/ARXIV.2207.10551}
\showDOI{\tempurl}


\bibitem[Thoppilan et~al\mbox{.}(2022)]%
        {lamda}
\bibfield{author}{\bibinfo{person}{Romal Thoppilan}, \bibinfo{person}{Daniel~De
  Freitas}, \bibinfo{person}{Jamie Hall}, \bibinfo{person}{Noam Shazeer},
  \bibinfo{person}{Apoorv Kulshreshtha}, \bibinfo{person}{Heng-Tze Cheng},
  \bibinfo{person}{Alicia Jin}, \bibinfo{person}{Taylor Bos},
  \bibinfo{person}{Leslie Baker}, \bibinfo{person}{Yu Du},
  \bibinfo{person}{YaGuang Li}, \bibinfo{person}{Hongrae Lee},
  \bibinfo{person}{Huaixiu~Steven Zheng}, \bibinfo{person}{Amin Ghafouri},
  \bibinfo{person}{Marcelo Menegali}, \bibinfo{person}{Yanping Huang},
  \bibinfo{person}{Maxim Krikun}, \bibinfo{person}{Dmitry Lepikhin},
  \bibinfo{person}{James Qin}, \bibinfo{person}{Dehao Chen},
  \bibinfo{person}{Yuanzhong Xu}, \bibinfo{person}{Zhifeng Chen},
  \bibinfo{person}{Adam Roberts}, \bibinfo{person}{Maarten Bosma},
  \bibinfo{person}{Vincent Zhao}, \bibinfo{person}{Yanqi Zhou},
  \bibinfo{person}{Chung-Ching Chang}, \bibinfo{person}{Igor Krivokon},
  \bibinfo{person}{Will Rusch}, \bibinfo{person}{Marc Pickett},
  \bibinfo{person}{Pranesh Srinivasan}, \bibinfo{person}{Laichee Man},
  \bibinfo{person}{Kathleen Meier-Hellstern}, \bibinfo{person}{Meredith~Ringel
  Morris}, \bibinfo{person}{Tulsee Doshi}, \bibinfo{person}{Renelito~Delos
  Santos}, \bibinfo{person}{Toju Duke}, \bibinfo{person}{Johnny Soraker},
  \bibinfo{person}{Ben Zevenbergen}, \bibinfo{person}{Vinodkumar Prabhakaran},
  \bibinfo{person}{Mark Diaz}, \bibinfo{person}{Ben Hutchinson},
  \bibinfo{person}{Kristen Olson}, \bibinfo{person}{Alejandra Molina},
  \bibinfo{person}{Erin Hoffman-John}, \bibinfo{person}{Josh Lee},
  \bibinfo{person}{Lora Aroyo}, \bibinfo{person}{Ravi Rajakumar},
  \bibinfo{person}{Alena Butryna}, \bibinfo{person}{Matthew Lamm},
  \bibinfo{person}{Viktoriya Kuzmina}, \bibinfo{person}{Joe Fenton},
  \bibinfo{person}{Aaron Cohen}, \bibinfo{person}{Rachel Bernstein},
  \bibinfo{person}{Ray Kurzweil}, \bibinfo{person}{Blaise Aguera-Arcas},
  \bibinfo{person}{Claire Cui}, \bibinfo{person}{Marian Croak},
  \bibinfo{person}{Ed Chi}, {and} \bibinfo{person}{Quoc Le}.}
  \bibinfo{year}{2022}\natexlab{}.
\newblock \bibinfo{title}{LaMDA: Language Models for Dialog Applications}.
\newblock
\newblock
\showeprint[arxiv]{2201.08239}~[cs.CL]


\bibitem[Touvron et~al\mbox{.}(2023)]%
        {llama}
\bibfield{author}{\bibinfo{person}{Hugo Touvron}, \bibinfo{person}{Thibaut
  Lavril}, \bibinfo{person}{Gautier Izacard}, \bibinfo{person}{Xavier
  Martinet}, \bibinfo{person}{Marie-Anne Lachaux}, \bibinfo{person}{Timothée
  Lacroix}, \bibinfo{person}{Baptiste Rozière}, \bibinfo{person}{Naman Goyal},
  \bibinfo{person}{Eric Hambro}, \bibinfo{person}{Faisal Azhar},
  \bibinfo{person}{Aurelien Rodriguez}, \bibinfo{person}{Armand Joulin},
  \bibinfo{person}{Edouard Grave}, {and} \bibinfo{person}{Guillaume Lample}.}
  \bibinfo{year}{2023}\natexlab{}.
\newblock \bibinfo{title}{LLaMA: Open and Efficient Foundation Language
  Models}.
\newblock
\newblock
\showeprint[arxiv]{2302.13971}~[cs.CL]


\bibitem[Trautmann et~al\mbox{.}(2022)]%
        {gpt-legal-promt_0}
\bibfield{author}{\bibinfo{person}{Dietrich Trautmann}, \bibinfo{person}{Alina
  Petrova}, {and} \bibinfo{person}{Frank Schilder}.}
  \bibinfo{year}{2022}\natexlab{}.
\newblock \showarticletitle{Legal Prompt Engineering for Multilingual Legal
  Judgement Prediction}.
\newblock \bibinfo{journal}{\emph{CoRR}}  \bibinfo{volume}{abs/2212.02199}
  (\bibinfo{year}{2022}).
\newblock
\urldef\tempurl%
\url{https://doi.org/10.48550/arXiv.2212.02199}
\showDOI{\tempurl}
\showeprint[arXiv]{2212.02199}


\bibitem[Tuggener et~al\mbox{.}(2020)]%
        {ledgar}
\bibfield{author}{\bibinfo{person}{Don Tuggener}, \bibinfo{person}{Pius von
  D{\"a}niken}, \bibinfo{person}{Thomas Peetz}, {and} \bibinfo{person}{Mark
  Cieliebak}.} \bibinfo{year}{2020}\natexlab{}.
\newblock \showarticletitle{{LEDGAR}: A Large-Scale Multi-label Corpus for Text
  Classification of Legal Provisions in Contracts}. In
  \bibinfo{booktitle}{\emph{Proceedings of the Twelfth Language Resources and
  Evaluation Conference}}. \bibinfo{publisher}{European Language Resources
  Association}, \bibinfo{address}{Marseille, France},
  \bibinfo{pages}{1235--1241}.
\newblock
\showISBNx{979-10-95546-34-4}
\urldef\tempurl%
\url{https://aclanthology.org/2020.lrec-1.155}
\showURL{%
\tempurl}


\bibitem[Vaswani et~al\mbox{.}(2017)]%
        {transformer}
\bibfield{author}{\bibinfo{person}{Ashish Vaswani}, \bibinfo{person}{Noam
  Shazeer}, \bibinfo{person}{Niki Parmar}, \bibinfo{person}{Jakob Uszkoreit},
  \bibinfo{person}{Llion Jones}, \bibinfo{person}{Aidan~N. Gomez},
  \bibinfo{person}{Lukasz Kaiser}, {and} \bibinfo{person}{Illia Polosukhin}.}
  \bibinfo{year}{2017}\natexlab{}.
\newblock \showarticletitle{Attention Is All You Need}.
\newblock \bibinfo{journal}{\emph{CoRR}}  \bibinfo{volume}{abs/1706.03762}
  (\bibinfo{year}{2017}).
\newblock
\showeprint[arXiv]{1706.03762}
\urldef\tempurl%
\url{http://arxiv.org/abs/1706.03762}
\showURL{%
\tempurl}


\bibitem[Wang and Komatsuzaki(2021)]%
        {gpt-j}
\bibfield{author}{\bibinfo{person}{Ben Wang} {and} \bibinfo{person}{Aran
  Komatsuzaki}.} \bibinfo{year}{2021}\natexlab{}.
\newblock \bibinfo{title}{{GPT-J-6B: A 6 Billion Parameter Autoregressive
  Language Model}}.
\newblock
  \bibinfo{howpublished}{\url{https://github.com/kingoflolz/mesh-transformer-jax}}.
\newblock


\bibitem[Xiao et~al\mbox{.}(2018)]%
        {Chaojun}
\bibfield{author}{\bibinfo{person}{Chaojun Xiao}, \bibinfo{person}{Haoxi
  Zhong}, \bibinfo{person}{Zhipeng Guo}, \bibinfo{person}{Cunchao Tu},
  \bibinfo{person}{Zhiyuan Liu}, \bibinfo{person}{Maosong Sun},
  \bibinfo{person}{Yansong Feng}, \bibinfo{person}{Xianpei Han},
  \bibinfo{person}{Zhen Hu}, \bibinfo{person}{Heng Wang}, {and}
  \bibinfo{person}{Jianfeng Xu}.} \bibinfo{year}{2018}\natexlab{}.
\newblock \showarticletitle{{CAIL2018:} {A} Large-Scale Legal Dataset for
  Judgment Prediction}.
\newblock \bibinfo{journal}{\emph{CoRR}}  \bibinfo{volume}{abs/1807.02478}
  (\bibinfo{year}{2018}).
\newblock
\showeprint[arXiv]{1807.02478}
\urldef\tempurl%
\url{http://arxiv.org/abs/1807.02478}
\showURL{%
\tempurl}


\bibitem[Xu et~al\mbox{.}(2020)]%
        {xu-etal}
\bibfield{author}{\bibinfo{person}{Nuo Xu}, \bibinfo{person}{Pinghui Wang},
  \bibinfo{person}{Long Chen}, \bibinfo{person}{Li Pan},
  \bibinfo{person}{Xiaoyan Wang}, {and} \bibinfo{person}{Junzhou Zhao}.}
  \bibinfo{year}{2020}\natexlab{}.
\newblock \showarticletitle{Distinguish Confusing Law Articles for Legal
  Judgment Prediction}. In \bibinfo{booktitle}{\emph{Proceedings of the 58th
  Annual Meeting of the Association for Computational Linguistics}}.
  \bibinfo{publisher}{Association for Computational Linguistics},
  \bibinfo{address}{Online}, \bibinfo{pages}{3086--3095}.
\newblock
\urldef\tempurl%
\url{https://doi.org/10.18653/v1/2020.acl-main.280}
\showDOI{\tempurl}


\bibitem[Yang et~al\mbox{.}(2016)]%
        {hierarchical-attention-networks}
\bibfield{author}{\bibinfo{person}{Zichao Yang}, \bibinfo{person}{Diyi Yang},
  \bibinfo{person}{Chris Dyer}, \bibinfo{person}{Xiaodong He},
  \bibinfo{person}{Alex Smola}, {and} \bibinfo{person}{Eduard Hovy}.}
  \bibinfo{year}{2016}\natexlab{}.
\newblock \showarticletitle{Hierarchical Attention Networks for Document
  Classification}. In \bibinfo{booktitle}{\emph{Proceedings of the 2016
  Conference of the North {A}merican Chapter of the Association for
  Computational Linguistics: Human Language Technologies}}.
  \bibinfo{publisher}{Association for Computational Linguistics},
  \bibinfo{address}{San Diego, California}, \bibinfo{pages}{1480--1489}.
\newblock
\urldef\tempurl%
\url{https://doi.org/10.18653/v1/N16-1174}
\showDOI{\tempurl}


\bibitem[Ye et~al\mbox{.}(2018)]%
        {ye-etal-interpretable}
\bibfield{author}{\bibinfo{person}{Hai Ye}, \bibinfo{person}{Xin Jiang},
  \bibinfo{person}{Zhunchen Luo}, {and} \bibinfo{person}{Wenhan Chao}.}
  \bibinfo{year}{2018}\natexlab{}.
\newblock \showarticletitle{Interpretable Charge Predictions for Criminal
  Cases: Learning to Generate Court Views from Fact Descriptions}. In
  \bibinfo{booktitle}{\emph{Proceedings of the 2018 Conference of the North
  {A}merican Chapter of the Association for Computational Linguistics: Human
  Language Technologies, Volume 1 (Long Papers)}}.
  \bibinfo{publisher}{Association for Computational Linguistics},
  \bibinfo{address}{New Orleans, Louisiana}, \bibinfo{pages}{1854--1864}.
\newblock
\urldef\tempurl%
\url{https://doi.org/10.18653/v1/N18-1168}
\showDOI{\tempurl}


\bibitem[Yu et~al\mbox{.}(2022)]%
        {gpt-legal-promt_1}
\bibfield{author}{\bibinfo{person}{Fangyi Yu}, \bibinfo{person}{Lee Quartey},
  {and} \bibinfo{person}{Frank Schilder}.} \bibinfo{year}{2022}\natexlab{}.
\newblock \bibinfo{title}{Legal Prompting: Teaching a Language Model to Think
  Like a Lawyer}.
\newblock
\newblock
\showeprint[arxiv]{2212.01326}~[cs.CL]


\bibitem[Zaheer et~al\mbox{.}(2020)]%
        {bigbird}
\bibfield{author}{\bibinfo{person}{Manzil Zaheer}, \bibinfo{person}{Guru
  Guruganesh}, \bibinfo{person}{Kumar~Avinava Dubey}, \bibinfo{person}{Joshua
  Ainslie}, \bibinfo{person}{Chris Alberti}, \bibinfo{person}{Santiago
  Ontanon}, \bibinfo{person}{Philip Pham}, \bibinfo{person}{Anirudh Ravula},
  \bibinfo{person}{Qifan Wang}, \bibinfo{person}{Li Yang}, {and}
  \bibinfo{person}{Amr Ahmed}.} \bibinfo{year}{2020}\natexlab{}.
\newblock \showarticletitle{Big Bird: Transformers for Longer Sequences}. In
  \bibinfo{booktitle}{\emph{Advances in Neural Information Processing
  Systems}}, \bibfield{editor}{\bibinfo{person}{H.~Larochelle},
  \bibinfo{person}{M.~Ranzato}, \bibinfo{person}{R.~Hadsell},
  \bibinfo{person}{M.F. Balcan}, {and} \bibinfo{person}{H.~Lin}} (Eds.),
  Vol.~\bibinfo{volume}{33}. \bibinfo{publisher}{Curran Associates, Inc.},
  \bibinfo{pages}{17283--17297}.
\newblock
\urldef\tempurl%
\url{https://proceedings.neurips.cc/paper_files/paper/2020/file/c8512d142a2d849725f31a9a7a361ab9-Paper.pdf}
\showURL{%
\tempurl}


\bibitem[Zeiler and Fergus(2014)]%
        {Zeiler}
\bibfield{author}{\bibinfo{person}{Matthew~D. Zeiler} {and}
  \bibinfo{person}{Rob Fergus}.} \bibinfo{year}{2014}\natexlab{}.
\newblock \showarticletitle{Visualizing and Understanding Convolutional
  Networks}. In \bibinfo{booktitle}{\emph{Computer Vision -- ECCV 2014}},
  \bibfield{editor}{\bibinfo{person}{David Fleet}, \bibinfo{person}{Tomas
  Pajdla}, \bibinfo{person}{Bernt Schiele}, {and} \bibinfo{person}{Tinne
  Tuytelaars}} (Eds.). \bibinfo{publisher}{Springer International Publishing},
  \bibinfo{address}{Cham}, \bibinfo{pages}{818--833}.
\newblock
\showISBNx{978-3-319-10590-1}


\bibitem[Zhang et~al\mbox{.}(2019)]%
        {hibert}
\bibfield{author}{\bibinfo{person}{Xingxing Zhang}, \bibinfo{person}{Furu Wei},
  {and} \bibinfo{person}{Ming Zhou}.} \bibinfo{year}{2019}\natexlab{}.
\newblock \showarticletitle{{HIBERT}: Document Level Pre-training of
  Hierarchical Bidirectional Transformers for Document Summarization}. In
  \bibinfo{booktitle}{\emph{Proceedings of the 57th Annual Meeting of the
  Association for Computational Linguistics}}. \bibinfo{publisher}{Association
  for Computational Linguistics}, \bibinfo{address}{Florence, Italy},
  \bibinfo{pages}{5059--5069}.
\newblock
\urldef\tempurl%
\url{https://doi.org/10.18653/v1/P19-1499}
\showDOI{\tempurl}


\bibitem[Zheng et~al\mbox{.}(2021)]%
        {case-hold}
\bibfield{author}{\bibinfo{person}{Lucia Zheng}, \bibinfo{person}{Neel Guha},
  \bibinfo{person}{Brandon~R. Anderson}, \bibinfo{person}{Peter Henderson},
  {and} \bibinfo{person}{Daniel~E. Ho}.} \bibinfo{year}{2021}\natexlab{}.
\newblock \showarticletitle{When Does Pretraining Help? Assessing
  Self-Supervised Learning for Law and the CaseHOLD Dataset of 53,000+ Legal
  Holdings}. In \bibinfo{booktitle}{\emph{Proceedings of the Eighteenth
  International Conference on Artificial Intelligence and Law}} (S\~{a}o Paulo,
  Brazil) \emph{(\bibinfo{series}{ICAIL '21})}. \bibinfo{publisher}{Association
  for Computing Machinery}, \bibinfo{address}{New York, NY, USA},
  \bibinfo{pages}{159–168}.
\newblock
\showISBNx{9781450385268}
\urldef\tempurl%
\url{https://doi.org/10.1145/3462757.3466088}
\showDOI{\tempurl}


\bibitem[Zhong et~al\mbox{.}(2018)]%
        {zhong-etal-2018-legal}
\bibfield{author}{\bibinfo{person}{Haoxi Zhong}, \bibinfo{person}{Zhipeng Guo},
  \bibinfo{person}{Cunchao Tu}, \bibinfo{person}{Chaojun Xiao},
  \bibinfo{person}{Zhiyuan Liu}, {and} \bibinfo{person}{Maosong Sun}.}
  \bibinfo{year}{2018}\natexlab{}.
\newblock \showarticletitle{Legal Judgment Prediction via Topological
  Learning}. In \bibinfo{booktitle}{\emph{Proceedings of the 2018 Conference on
  Empirical Methods in Natural Language Processing}}.
  \bibinfo{publisher}{Association for Computational Linguistics},
  \bibinfo{address}{Brussels, Belgium}, \bibinfo{pages}{3540--3549}.
\newblock
\urldef\tempurl%
\url{https://doi.org/10.18653/v1/D18-1390}
\showDOI{\tempurl}


\bibitem[Zhong et~al\mbox{.}(2020a)]%
        {Zhong_Wang_2020}
\bibfield{author}{\bibinfo{person}{Haoxi Zhong}, \bibinfo{person}{Yuzhong
  Wang}, \bibinfo{person}{Cunchao Tu}, \bibinfo{person}{Tianyang Zhang},
  \bibinfo{person}{Zhiyuan Liu}, {and} \bibinfo{person}{Maosong Sun}.}
  \bibinfo{year}{2020}\natexlab{a}.
\newblock \showarticletitle{Iteratively Questioning and Answering for
  Interpretable Legal Judgment Prediction}.
\newblock \bibinfo{journal}{\emph{Proceedings of the AAAI Conference on
  Artificial Intelligence}} \bibinfo{volume}{34}, \bibinfo{number}{01}
  (\bibinfo{date}{Apr.} \bibinfo{year}{2020}), \bibinfo{pages}{1250--1257}.
\newblock
\urldef\tempurl%
\url{https://doi.org/10.1609/aaai.v34i01.5479}
\showDOI{\tempurl}


\bibitem[Zhong et~al\mbox{.}(2020b)]%
        {diff_from_generalTxt_1}
\bibfield{author}{\bibinfo{person}{Haoxi Zhong}, \bibinfo{person}{Chaojun
  Xiao}, \bibinfo{person}{Cunchao Tu}, \bibinfo{person}{Tianyang Zhang},
  \bibinfo{person}{Zhiyuan Liu}, {and} \bibinfo{person}{Maosong Sun}.}
  \bibinfo{year}{2020}\natexlab{b}.
\newblock \showarticletitle{How Does {NLP} Benefit Legal System: A Summary of
  Legal Artificial Intelligence}. In \bibinfo{booktitle}{\emph{Proceedings of
  the 58th Annual Meeting of the Association for Computational Linguistics}}.
  \bibinfo{publisher}{Association for Computational Linguistics},
  \bibinfo{address}{Online}, \bibinfo{pages}{5218--5230}.
\newblock
\urldef\tempurl%
\url{https://doi.org/10.18653/v1/2020.acl-main.466}
\showDOI{\tempurl}


\end{thebibliography}


\end{document}